\documentclass[preprint,showpacs,preprintnumbers,amsmath,amssymb]{revtex4}

\usepackage{graphicx,epsfig}
\usepackage{dcolumn}
\usepackage{bm}

\begin{document}

\preprint{}

\title{Effects of Diversity on Multi-agent Systems:\\
Minority Games
}

\author{K. Y. Michael Wong}
\author{S. W. Lim}
\author{Zhuo Gao}

\affiliation{%
Department of Physics, Hong Kong University of Science and Technology, 
Clear Water Bay, Hong Kong, China
}%

\date{\today}

\begin{abstract}
We consider a version of large population games whose agents compete for
resources using strategies with adaptable preferences. 
The games can be used to model economic markets, ecosystems 
or distributed control. 
Diversity of initial preferences of strategies is introduced by randomly 
assigning biases to the strategies of different agents. 
We find that diversity among the agents reduces their maladaptive behavior. 
We find interesting scaling relations with diversity 
for the variance and other parameters such as the convergence time, 
the fraction of fickle agents, and the variance of wealth, illustrating 
their dynamical origin. 
When diversity increases, the scaling dynamics is modified 
by kinetic sampling and waiting effects. 
Analyses yield excellent agreement with simulations.
\end{abstract}

\maketitle


\section{\label{sec:intro}Introduction}

Many natural and artificial systems involve interacting agents, each making
independent decisions to compete for limited resources, but globally exhibit
coordinated behavior through their mutual adaptation 
\cite{global,weiss,schweitzer,minority}. 
Examples include
the formation of ecological patterns 
due to the competition of predators hunting
for food, the price adjustment due to the competition of buyers or sellers in
economic markets, and the load adjustment due to the competition of distributed
controllers of packet flows in computer networks. While a standard approach is
to analyse the steady state behavior of the system described by the Nash
equilibria \cite{game}, 
it is legitimate to consider how the steady state is approached,
since such processes are dynamical in nature, 
and the approach may be interfered
by the presence of periodic, chaotic or metastable attractors. 
Dynamical studies
are especially relevant when one considers the effects of changing environment,
such as that in economics or distributed control.

The recently proposed Minority Games (MG) are prototypes of such multi-agent
systems \cite{minority}. 
Extensive studies have revealed the steady-state properties of the
game when the complexity of the agents is high \cite{challet}. 
On the other hand, the
dynamical nature of the adaptive processes is revealed when the complexity of
the agents is low, 
wherein the final states of the system depend on the initial conditions, 
and the system often ends up with large fluctuations at final states,
much remote from the efficient state predicted by equilibrium studies 
\cite{challet,garrahan}.   
The large fluctuations in the original MG is related to the uniformly zero   
preference of strategies for all agents. This has to be re-examined 
for at least two reasons. 
First, when the game is used to model economic systems, 
it is not realistic to expect that all agents have the same preference 
when they enter the market. 
Rather, the agents have their own preferences 
according to their individual objectives, expectations and available capital. 
For example, some have stronger inclinations towards aggressive strategies, 
and others more conservative. 
Furthermore, in games which use public information only, identical  
initial preferences imply that different agents would maintain identical
preferences of strategies at all subsequent steps of the game, 
which is again unlikely. 
Second, when the game is used to model distributed control 
in multi-agent systems, 
identical preferences of strategies of the agents lead to
{\it maladaptive} behavior, 
which refers to the bursts of the population's decisions
due to the agents' premature rush to certain state \cite{savit,johnson}.
As a result, the population difference 
between the majority and minority groups is large.
For economic markets, this corresponds to large price fluctuations; 
for distributed control, this corresponds to an uneven resource allocation; 
both imply low system efficiency.
Hence, maladaptation hinders the attainment of optimal system efficiency.

There have been many attempts to improve the system efficiency. 
For example,   
thermal noise \cite{cavagna} or biased strategies \cite{yip} 
are found to reduce the fluctuations. 
More relevant to this work, there were indications that
maladaptation can be reduced by appropriate choices of the initial condition 
at the low complexity phase. 
The dependence of initial conditions was noted in the  
replica approach to the exogenous MG \cite{challet}. 
System efficiency can be improved
by random initial conditions in the original MG \cite{moro}, 
or systems driven by 
vectorized external information \cite{garrahan}. 
It was noted that the reduced variance can be obtained hysteretically 
by quasistatic increase and decrease of
the complexity from an unbiased initial condition, clearly demonstrating the
non-equilibrium nature of this phenomenon \cite{sherrington}. 
By generalizing the
strategy evaluation mechanism to the batch mode, and using a payoff function
linear in the winning margin, 
the generating functional analysis showed that    
fluctuations are reduced by biased starts of the agents' strategy payoff
valuations \cite{heimel}. 
The same is valid in its noisy extension \cite{coolen}. However,
no systematic studies about the effects of random biases have been made.

In this paper, we consider the effects of randomness in the initial
preferences of strategies among the agents. 
Initial conditions can be selected
to make the system dynamics completely deterministic, 
thus yielding highly precise simulation results 
useful for refined comparison with theories. 
As we shall see, 
a consequence of this {\it diversity} 
is that agents sharing common strategies 
are less likely to adopt them at the same time, 
and maladaptation is reduced. 
This results in an improved system efficiency, 
as reflected by the reduced variance of the population decisions. 
We find interesting scaling
relations with the diversity for the variance, and a number of dynamical
parameters, such as the convergence time, the fraction of fickle agents, 
and the
variance of wealth, illustrating their dynamical origin. 
When diversity increases, 
we find that the scaling dynamics is modified by a sampling mechanism
self-imposed by the requirement of the dynamics to stay in the attractor, an   
effect we term {\it kinetic sampling}. 
Preliminary results have been sketched in \cite{wong}.

This paper is organized as follows. 
After introducing the Minority Game in      
Section~\ref{sec:mg}, we discuss the variation of fluctuations
when diversity increases,
identifying 3 regimes of behavior: 
multinomial, scaling, and kinetic sampling,  
analyzed in Sections~\ref{sec:multi} to \ref{sec:kinetic} respectively.
Besides the fluctuations, other dynamical properties,
namely, the fraction of fickle agents, the convergence time,
and the variance of wealth,
are discussed in Sections~\ref{sec:fickle} to \ref{sec:wealth} respectively.
The paper is concluded in Section~\ref{sec:discuss}.

\section{\label{sec:mg}The Minority Game}

We consider a population of $N$ agents competing selfishly 
to be in the minority group in an environment of limited resources,
$N$ being odd \cite{minority}.
Each of the $N$ agents can make a decision 1 or 0 at each time step,
and the minority group wins.
For typical control tasks such as the distribution of shared resources,
the decisions 1 and 0 may represent two atternative resources,
so that less agents utilizing a resource implies more abundance.
For economic markets,
the decisions 1 and 0 correspond to buying and selling respectively,
so that the buyers can win by belonging to the minority group,
as a consequence of the price being pushed down 
when supply is greater than demand, and vice versa.

Each agent makes her decision independently
according to her own finite set of strategies,                
randomly picked before the game starts.
Each of her $s$ strategies is based on the history of the game,
which is the time series of the winning bits in the most recent $m$ steps.
Hence, $m$ is the memory size. 
There are $D\equiv 2^m$ possible histories,
thus $D$ is the dimension of the strategy space.
While most previous work considered the case $D\sim N$,
we will mainly study the case $m\gtrsim 1$ in this paper.
As we shall see,
this simplification enables us to make detailed analysis of the system,
revealing many new features.

A strategy is then a Boolean function
which maps each of the $D$ histories to decisions 1 or 0.
Denoting the winning state at time $t$ by $\sigma(t)$ ($\sigma(t)=1, 0$), 
we can convert an $m$-bit history
$\sigma(t-m+1),\cdots,\sigma(t)$
to an integer {\it historical state} $\mu^*(t)$ of modulo $D$, given by
\begin{equation}
	\mu^*(t)=\sum_{t'=0}^{m-1}\sigma(t-t')2^{t'},
\end{equation}
and the Boolean decisions of strategy $a$ responding to input state $\mu$ 
are denoted by $\sigma_a^\mu=1, 0$, 
corresponding to the binary decisions $\xi_a^\mu=\pm 1$
via $\xi_a^\mu\equiv 2\sigma_a^\mu-1$.
For subsequent analyses of strategies,
the label $a$ of a strategy is given by an integer between 0 and $2^D-1$,
where
\begin{equation}
        a=\sum_{\mu=0}^{D-1}\sigma_a^\mu 2^{D-1-\mu}.
\label{stlabel}
\end{equation}

The success of a strategy is measured by its {\it cumulative payoff}
(also called {\it virtual point} in the literature),
which increases (decreases) by $1$
if it indicates a winning (losing) decision at a time step.
Note that the payoffs attributed to the strategies at each step 
depend only on the signs of the decisions, 
and is independent of the magnitude of the winning margins. 
This is called the {\it step payoff}, 
and follows the original version of the MG \cite{minority}. 
Many recent studies used payoffs 
with magnitudes increasing with the difference 
between the majority and minority population. 
In particular, payoffs that are linear in the population difference 
are called {\it linear payoffs}, 
and are found convenient in the application of analytical techniques 
such as the replica method \cite{challet} 
or the generating functional analysis \cite{heimel}. 
In the analysis of this paper, the step payoff is more convenient.

At each time step, 
each agent chooses, out of her $s$ strategies, 
the one with the highest cumulative payoff
(updated every step irrespective of whether it is adopted or not) 
and makes decisions accordingly.
The difference between the total number of winning and losing decisions 
of an agent up to a time step is called
her {\it wealth} at that time.
The long-term goal of an agent is to maximize her wealth. 

To model diversity among the agents,
the agents may enter the game with diverse preferences of their strategies.
This means that each agent has random integer {\it biases} 
to the initial cumulative payoffs of each of her $s$ strategies. 
We are interested in how the extent of randomness 
affects the system behavior, 
and there are many choices of the bias distribution. 
A natural choice is the multinomial distribution, 
which can be modeled by assigning integer biases 
to the $s$ strategies of each agent,
which add up to an odd integer $R$.
Then, the biased payoff of a strategy of an agent 
obeys a multinomial distribution with mean $R/s$ and variance $R(s-1)/s^2$.
The ratio $\rho\equiv R/N$ is referred to as the {\it diversity}.

For the binomial case $s=2$ and odd $R$, which will be studied here,
no two strategies have the same cumulative payoffs throughout the game.
Hence there are no ties,
and the dynamics of the game is deterministic,
resulting in highly precise simulation results
useful for refined comparison with theories.
This is in contrast with previous versions of the game,
which correspond to the special case of $R=0$.

Furthermore, for an agent holding strategies $a$ and $b$ (with $a < b$),
the biases affect her decisions only through the bias difference $\omega$ 
of strategy $a$ with respect to $b$.
Hence we let $S_{ab}(\omega)$ be the number of agents 
holding strategies $a$ and $b$,
where the bias of strategy $a$ is displaced by $\omega$ with respect to $b$,
and its disordered average is 
\begin{equation}
	\langle S_{ab}(\omega)\rangle
	=\frac{N}{2^{2D-1}}\frac{1}{2^R}\binom{R}{\frac{R-\omega}{2}}.
\label{average}
\end{equation}
To describe the macroscopic dynamics of the system,
we define the $D$-dimensional phase space with the components $A^\mu(t)$,
which is the fraction of agents making decision 1   
responding to input $\mu$ of their used strategies,
subtracted by that for decision $0$.
While only one of the $D$ components corresponds to the historical state
$\mu^*(t)$ of the system,
the augmentation to $D$ components is necessary to describe
the attractor structure and the transient behavior of the system dynamics.

The key to analysing the system dynamics 
is the observation that the cumulative payoffs of all strategies
displace by exactly the same amount
when the game proceeds,
though their initial values may be different.
Hence for a given strategy pair,
the profile of the cumulative payoff distribution remains binomial,
but the peak position shifts with the game dynamics.
Hence once the cumulative payoffs are known,
the state location in the $D$-dimensional phase space is given by
\begin{eqnarray}
	A^\mu(t)&=&\frac{1}{N}\sum_{a<b,\omega}S_{ab}(\omega)
	\left\{\Theta[\omega+\Omega_a(t)-\Omega_b(t)]\xi_a^\mu
	+\Theta[-\omega-\Omega_a(t)+\Omega_b(t)]\xi_b^\mu\right\}
	\nonumber\\
	& &+\frac{1}{N}\sum_a S_a\xi_a^\mu,
\label{amu}
\end{eqnarray}
where $\Omega_a(t)$ is the cumulative payoff of strategy $a$ at time $t$,
$S_a$ is the number of agents holding 2 identical strategies labelled $a$,
and $\Theta(x)$ is the step function of $x$.
For agents holding non-identical strategies $a < b$,
the agents make decision according to strategy $a$ if
$\omega+\Omega_a(t)-\Omega_b(t)>0$, and strategy $b$ otherwise.
Hence $\omega+\Omega_a(t)-\Omega_b(t)$ 
is referred to as the {\it preference} of $a$ with respect to $b$.
In turn, the cumulative payoff of a strategy $a$ is updated by
\begin{equation}
	\Omega_a(t+1)=\Omega_a(t)-\xi_a^{\mu^*(t)}{\rm sgn}A^{\mu^*(t)}(t).
\label{cumpay}
\end{equation}

\begin{figure}
\centerline{\epsfig{figure=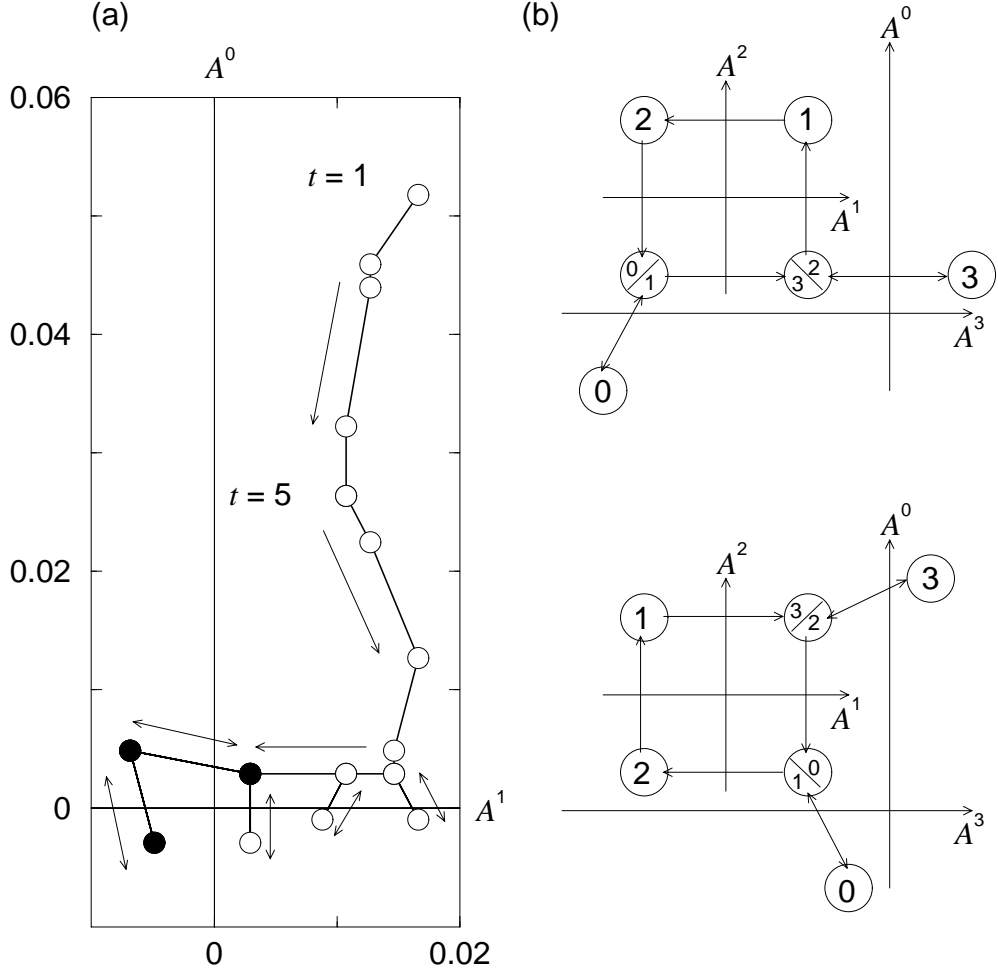,width=0.8\linewidth}}
\caption{\label{phase}
(a) The state motion of a sample in the phase space for $m=1$, 
$s=2$, $N=1023$ and $R=16383$.
Empty dots: transient states.
Solid dots: attractor states.
(b) The attractors in the phase subspace of $A^1$ and $A^2$ for $m=2$.
6 of the 8 states remain in the second quadrant of the subspace 
formed by $A^3$ and $A^0$. 
The location of the other 2 states are indicated 
in the $A^3$ and $A^0$ subspace, 
instead of the $A^1$ and $A^2$ subspace.
The numbers in the circles denote the elements of the attractor sequences
in Eqs.~(\ref{m2first}) and (\ref{m2second}).}
\end{figure}

Fig.~\ref{phase}(a) illustrates the convergence to the attractor
for the visualizable case of $m=1$.
The dynamics proceeds in the direction
which tends to reduce the magnitude of the components of $A^\mu(t)$
\cite{challet}.
However, a certain amount of maladaptation always exists in the system,
so that the components of $A^\mu(t)$ overshoot,
resulting in periodic attractors of period $2D$,
as reported in the literature \cite{zheng, lee}.
The state evolution is given by the integer equation
\begin{equation}
	\mu^*(t+1)={\rm mod}(2\mu^*(t)+\sigma(t),D),
\label{evolution}
\end{equation}
so that every state $\mu$ appears as historical states
two times in a steady-state period,
with $\sigma(t)$ appearing as 0 and 1, each exactly once.
One occurence brings $A^\mu$ from positive to negative,
and another bringing it back from negative to positive,
thus completing a cycle. 
The components keep on oscillating,
but never reach zero.
This results in an {\it antipersistent} time series \cite{metzler}.
For the example in Fig.~\ref{phase}(a),
the steady state is described by the sequence 
\begin{equation}
	\mu(t)=\sigma(t)=0,1,1,0,
\label{m1}
\end{equation}
where one notes that both states 0 and 1 are followed by 0 and 1 once each.

For $m=2$, there are 2 attractor sequences as shown in Fig.~\ref{phase}(b),
\begin{equation}
	\mu(t)=0,1,3,3,2,1,2,0,
\label{m2first}
\end{equation}
and
\begin{equation}
	\mu(t)=0,1,2,1,3,3,2,0.
\label{m2second}
\end{equation}
Again, one notes that each of the states 0, 1, 2, 3
are followed by an even ($\sigma(t)=0$)
and an odd state ($\sigma(t)=1$) once each.
Furthermore, we note that the attractor sequences 
in Eqs.~(\ref{m2first}) and (\ref{m2second})
are related by the conjugation symmetry $\mu(t)\rightarrow 3-\mu(t)$.
For general values of $m$,
an attractor sequence can be obtained 
by starting with the state $\mu^*(0)=\sigma(0)=0$,
and assigning $\sigma(t)=1$ 
if the value of $\mu^*(t)$ appears the first time in the sequence,
and 0 the second time, 
such as the attracters in Eq.~(\ref{m1}) and (\ref{m2first}).
In general, other attractor sequences can be obtained by computer search,
and the number of attractor sequences can be verified to be $2^D/2D$,
which forms the de Bruijn sequence in terms of $m$,
corresponding to the number of distinct ring configurations of length $2D$,
for which all sub-strings of length $m+1$ are distinct \cite{deBruijn}.

The population averages of the decisions 
oscillate around 0 at the steady state. 
Since a large difference between the majority and minority populations 
implies inefficient resource allocation, 
the inefficiency of the game is often characterized by 
the normalized variance $\sigma^2/N$ of the population making decision 1 
at the steady state.
Since this population size at time $t$ is given by $N(1-A^{\mu^*(t)})/2$, 
we have
\begin{equation}\label{var}
	\frac{\sigma^2}{N}
	=\lim_{t\rightarrow\infty}\frac{N}{4}
	\langle[A^{\mu^*}(t)-\langle A^{\mu^*}(t)\rangle_t]^2\rangle_t,
\end{equation}
where $\langle\quad\rangle_t$ denotes time average at the steady state.

\begin{figure}
\centerline{\epsfig{figure=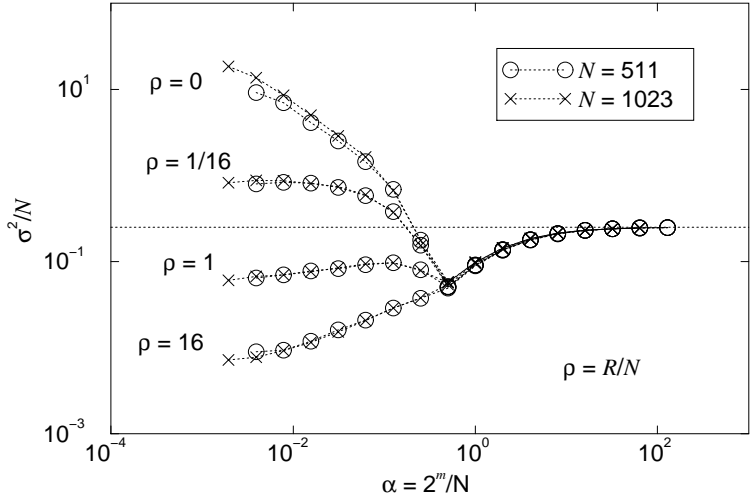,width=0.6\linewidth}}
\caption{\label{complexity}
The dependence of the variance of the population making decision $+$ 
on the complexity for different diversities at $s=2$
averaged over 128 samples. 
The horizontal dotted line is the limit of random decisions.}
\end{figure}

As shown in Fig.~\ref{complexity},
the variance $\sigma^2/N$ of the population for decision 1
scales as a function of the {\it complexity} $\alpha\equiv D/N$,
agreeing with previous observations \cite{savit}.
When $\alpha$ is small, games with increasing complexity
create time series of decreasing fluctuations.
A phase transition takes place around $\alpha_c\approx 0.3$,
after which it increases gradually to the limit of random decisions,
with $\sigma^2/N=0.25$.
When $\alpha <\alpha_c$, the occurences of decision 1 and 0
responding to a given historical state $\mu$ are equal,
and is referred to as the {\it symmetric} phase \cite{symmetry}.
On the other hand, in the {\it asymmetric} phase above $\alpha_c$,
the occurences of decisions are biased for at least some history $\mu$.

Figure~\ref{complexity} also shows the data collapse of the variance
for different values of diversity $\rho$.
It is observed that the variance decreases significantly with diversity
in the symmetric phase,
and remains unaffected in the asymmetric phase \cite{compare1}.
Furthermore, for a game efficiency prescribed by a given variance $\sigma^2/N$,
the required complexity of the agents is much reduced.

\begin{figure}
\centerline{\epsfig{figure=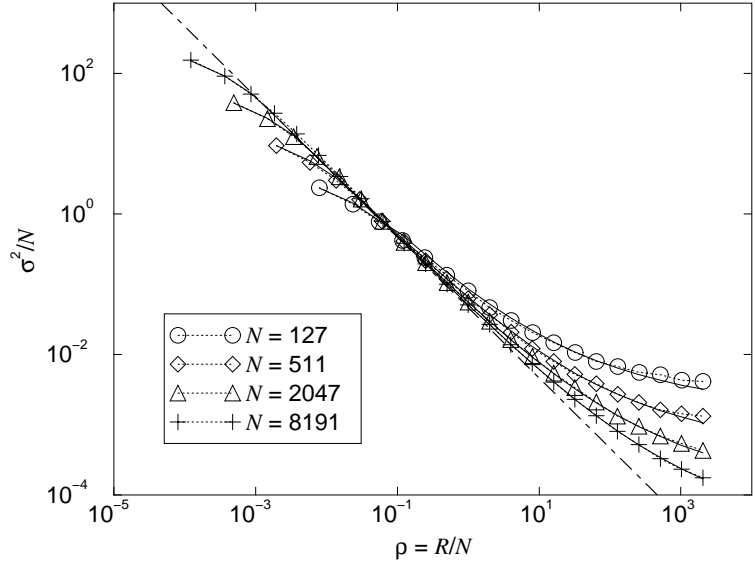,width=0.6\linewidth}}
\caption{\label{m1diversity}
The dependence of the variance of the population making decision $+$ 
on the diversity at $m=1$ and $s=2$.
Symbols: simulation results averaged over 1024 samples. 
Solid lines: theory.
Dashed-dotted line: scaling prediction.}
\end{figure}

\begin{figure}
\centerline{\epsfig{figure=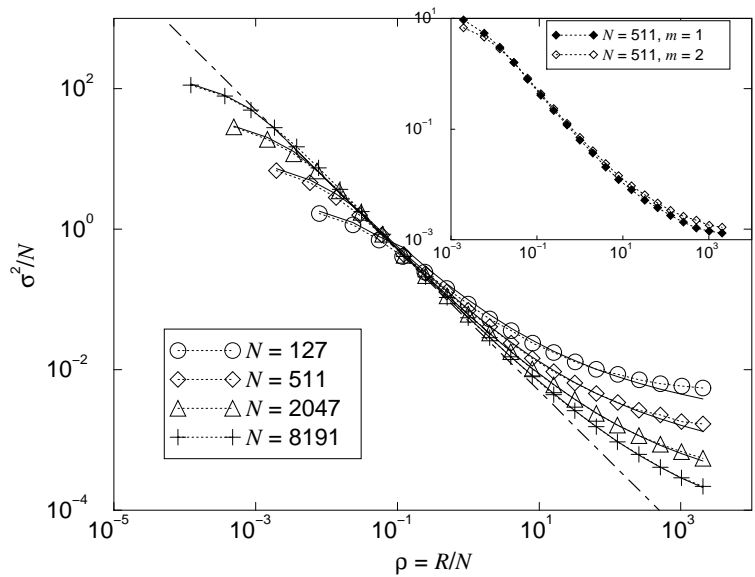,width=0.6\linewidth}}
\caption{\label{m2diversity}
The dependence of the variance of the population making decision $+$ 
on the diversity at $m=2$ and $s=2$.
Notations are the same as those of Fig.~\ref{m1diversity}.
Inset: A comparison of the variances at $m=1$ and $m=2$ 
in Figs.~\ref{m1diversity} and \ref{m2diversity}.}
\end{figure}

The dependence of the variance on the diversity is further shown
in Figs.~\ref{m1diversity} and \ref{m2diversity} 
for memory sizes $m=1$ and $m=2$ respectively.
The following three regimes can be identified 
and explained in Sections~\ref{sec:multi} to \ref{sec:kinetic} respectively:
(a) {\it multinomial regime:} when $\rho\sim N^{-1}$,
$\sigma^2/N\sim N$ with proportionality constants dependent on $m$;
(b) {\it scaling regime:} when $\rho\sim 1$,
$\sigma^2/N\sim \rho^{-1}$ with proportionality constants
independent of $m$ for $m$ not too large;
(c) {\it kinetic sampling regime:} when $\rho\sim N$,
$\sigma^2/N$ deviates above the scaling with $\rho^{-1}$
due to kinetic sampling effects as explained below,
and the scaling is given by $\sigma^2/N\sim f_m(\Delta)/N$,
where $\Delta$ is the {\it kinetic step size} given by 
\begin{equation}
        \Delta\equiv N\sqrt{\frac{2}{\pi R}}=\sqrt{\frac{2N}{\pi\rho}},
\label{stepdef}
\end{equation}
and $f_m$ is a function dependent on the memory size $m$.

To analyse the behavior in these regimes,
we derive the following expression 
for the step $\Delta A^{\mu}(t)\equiv A^{\mu}(t+1) - A^{\mu}(t)$,
at time $t$. Using Eq.~(\ref{amu}), we have 
\begin{equation}
	\Delta A^\mu(t)=\frac{1}{N}\sum_{a<b,\omega}S_{ab}(\omega)  
        \{\Theta[\omega+\Omega_a(t+1)-\Omega_b(t+1)]
        -\Theta[\omega+\Omega_a(t)-\Omega_b(t)]\}
        (\xi_a^\mu-\xi_b^\mu).
\label{step}
\end{equation}
Since the arguments of the step functions are odd integers,
nonzero contributions to Eq.~(\ref{step}) come from terms with
$\omega+\Omega_a(t+1)-\Omega_b(t+1)=\pm 1$ and 
$\omega+\Omega_a(t)-\Omega_b(t)=\mp 1$.
Using Eq.~(\ref{cumpay}),
the two arguments differ by $-(\xi_a^\mu-\xi_b^\mu){\rm sgn}A^\mu(t)$ 
with $\mu=\mu^*(t)$.
Hence the conditions for nonzero contributions become equivalent to
$\omega+\Omega_a(t)-\Omega_b(t)=\mp 1$ and
$\xi_a^\mu-\xi_b^\mu=\mp2{\rm sgn}A^\mu(t)$
for $\mu=\mu^*(t)$.
This reduces the steps to
\begin{equation}
	\Delta A^\mu(t)=\frac{1}{N}\sum_{a<b,\omega,\pm}S_{ab}(\omega)  
        \delta(\omega+\Omega_a(t)-\Omega_b(t)\pm 1)
        \delta(\xi_a^\mu-\xi_b^\mu\pm2{\rm sgn}A^\mu(t))
	(\pm)(\xi_a^\mu-\xi_b^\mu),
\label{stepgeneral}
\end{equation}
where $\mu=\mu^*(t)$, and $\delta(n)=1$ if $n=0$, and 0 otherwise.
For $\mu=\mu^*(t)$, this can be further simplified to
\begin{equation}
        \Delta A^\mu(t)=-{\rm sgn}A^\mu(t)\frac{2}{N}\sum_{a<b,\pm}
	S_{ab}(\mp 1-\Omega_a(t)+\Omega_b(t))
        \delta(\xi_a^\mu-\xi_b^\mu\pm 2{\rm sgn}A^\mu(t)),
\label{stepchange}
\end{equation}
To interpret this result,
we note that changes in $A^\mu(t)$
are only contributed by {\it fickle} agents
with marginal preferences of their strategies.
That is, those with
$\omega+\Omega_\alpha(t)-\Omega_\beta(t)=\pm 1$
and
$\xi_\alpha^\mu-\xi_\beta^\mu=\mp2{\rm sgn}A^\mu(t)$ for $\mu=\mu^*(t)$.
Furthermore, the step points in the direction 
that reduces the magnitude of $A^\mu(t)$.

Similarly, the steps along the direction $\nu$ 
other than the historical state $\mu^*(t)$ are given by
\begin{equation}
	\Delta A^\nu(t)=\frac{1}{N}\sum_{a < b, \pm}
	S_{ab}(\mp1-\Omega_a(t)+\Omega_b(t))
	\delta(\xi_a^\mu-\xi_b^\mu\pm2{\rm sgn}A^\mu(t))
	(\pm)(\xi_a^\nu-\xi_b^\nu)
\end{equation}
where $\mu=\mu^*(t)$. 
This shows that the steps 
along the non-historical direction are contributed 
by the subset of those fickle agents 
that contribute to the step along the historical direction,
and they can be positive or negative.

Next we consider the disordered average of the steps 
in Eq.~(\ref{stepgeneral}).
For this purpose, it is convenient to decompose the cumulative payoffs as
\begin{equation}\label{decomp}
        \Omega_a(t)=\sum_\mu k_\mu(t)\xi_a^\mu,
\end{equation}
where $k_\mu(t)$ is the number of wins minus losses of decision 1 
up to time $t$
when the game responded to history $\mu$. 
Since there are $2^D$ variables of $\Omega_a(t)$ 
and $D$ variables of $k_\mu(t)$,
this decomposition greatly simplifies the analysis,
and describes explicity how $\Omega_a(t)$ depends on the strategy decisions.
Introducing the integral representation of the Kronecka delta 
for the preference,
we can factorize the contributions of $\Omega_a(t)-\Omega_b(t)$ 
into a product over the states,
\begin{equation}
	\delta(\omega+\Omega_a(t)-\Omega_b(t)\pm 1)=
	\int_{0}^{2\pi}\frac{d\theta}{2\pi}e^{i\theta(\omega\pm 1)}
	\prod_{\lambda}e^{i\theta k_\lambda(\xi_a^\lambda-\xi_b^\lambda)},
\end{equation}
where the explicit dependence on $t$ is omitted for convenience here 
and in the subsequent derivation.
Using the identities
\begin{eqnarray}
        \delta(\xi_a^\mu-\xi_b^\mu\pm2{\rm sgn}A^\mu(t))&=&
        \frac{1}{4}[1\mp(\xi_a^\mu-\xi_b^\mu){\rm sgn}A^\mu
        -\xi_a^\mu\xi_b^\mu],\\
        e^{i\phi(\xi_a^\mu-\xi_b^\mu)}
        &=&\cos^2\phi+(\xi_a^\mu-\xi_b^\mu)i\sin\phi\cos\phi
        +\xi_a^\mu\xi_b^\mu\sin^2\phi,
\end{eqnarray}
and introducing the average in Eq.~(\ref{average}),
we obtain the following factorized expression from Eq.~(\ref{stepgeneral}) 
for $\mu=\mu^*(t)$,
\begin{eqnarray}\label{steplong}
        \langle\Delta A^\mu(t)\rangle=
        &&\frac{1}{2^{2D-1}}\sum_{a<b,\omega,\pm}
        \binom{R}{\frac{R-\omega}{2}}\frac{1}{2^R}
        \int_{0}^{2\pi}\frac{d\theta}{2\pi}e^{i\theta(\omega\pm 1)}
        \nonumber\\
        &&\frac{1}{4}[1\mp(\xi_a^\mu-\xi_b^\mu){\rm sgn}A^\mu
        -\xi_a^\mu\xi_b^\mu](\pm)(\xi_a^\mu-\xi_b^\mu)\nonumber\\
        &&[\cos^2k_\mu\theta
        +(\xi_a^\mu-\xi_b^\mu)i\sin k_\mu\theta\cos k_\mu\theta
        +\xi_a^\mu\xi_b^\mu\sin^2 k_\mu\theta]\nonumber\\
        &&\prod_{\lambda\neq\mu}
        [\cos^2k_\lambda\theta
        +(\xi_a^\lambda-\xi_b^\lambda)
        i\sin k_\lambda\theta\cos k_\lambda\theta
        +\xi_a^\lambda\xi_b^\lambda\sin^2 k_\lambda\theta].
\end{eqnarray}
The summation over $a<b$ can now be replaced 
by half times the independent summations over $a$ and $b$.
Noting that for given states $\mu,\nu\ldots\lambda$,
\begin{equation}
        \sum_a\xi_a^\mu\xi_a^\nu\cdots\xi_a^\lambda=0,
\end{equation} 
we find that all terms in the expansion of Eq.~(\ref{steplong})
vanish if they contain unpaired decisions $\xi_a^\nu$ or $\xi_b^\nu$.
The final result is 
\begin{equation}
        \langle\Delta A^\mu(t)\rangle=-{\rm sgn}A^\mu
        \int_{0}^{2\pi}\frac{d\theta}{2\pi}
        \cos^R\theta\cos(2k_\mu-{\rm sgn}A^\mu)\theta
        \prod_{\nu\neq\mu}\cos^2k_\nu\theta.
\label{stephist}
\end{equation}
Eq.~(\ref{stephist}) describes the change induced by the payoff component 
$k_\mu(t)$ incremented by $-{\rm sgn}A^\mu(t)$.
Since the step size depends on time implicitly through the payoff components,
the sum of all changes induced by $k_\mu(t)$ incremented from 0 yields
\begin{equation}
	\langle A^\mu(t)-A^\mu(0)\rangle
	=\int_{0}^{2\pi}\frac{d\theta}{2\pi}\cos^R\theta
	\frac{\sin k_\mu\theta\cos k_\mu\theta}{\sin\theta}
	\prod_{\nu\neq\mu}\cos^2k_\nu\theta.
\label{stepsum}
\end{equation} 
Similarly, the steps along the non-historical direction are given by
\begin{equation}
        \langle\Delta A^\nu(t)\rangle
        =\int_{0}^{2\pi}\frac{d\theta}{2\pi}\cos^R\theta
        \sin k_\nu\theta\cos k_\nu\theta \sin(2k_\mu{\rm sgn}A^\mu-1)\theta
        \prod_{\lambda\neq\mu\nu}\cos^2k_\lambda\theta,
\label{stepnonh}
\end{equation}
where $\nu\neq\mu^*(t)=\mu$.
The same result can be obtained from Eq.~(\ref{stepsum})
by considering the difference of 2 equations 
when one of the states labeled $\nu$ become historical and
$k_\nu$ changes by $-{\rm sgn}A^\nu$.

\section{\label{sec:multi}The Multinomial Regime}

When $\rho\sim N^{-1}$, or $R\sim 1$, 
there is a finite number of clusters of agents 
who make identical decisions throughout the game.
Since there are many agents in a typical cluster, 
their identical decisions will cause large fluctuations in their behavior.
Consider the example of $m=1$ and $R=1$. 
There are only 4 strategies.
For a pair of distinct strategies,
there is an average of $N/8$ agents picking them,
and $N/16$ agents in each cluster with biases $\pm 1$.
As a result,
we have $\sigma^2/N\sim N$.
The proportionality constant depends on $m$,
and is sensitive to the profile of the bias distribution.
Since we consider the multinomial distribution in Eq.~(\ref{average}) 
in this paper,
we call this the {\it multinomial} regime.
Another choice in the literature is the bimodal distribution 
\cite{moro,garrahan,sherrington,heimel,coolen},
which may have different behavior.

\begin{figure}
\centerline{\epsfig{figure=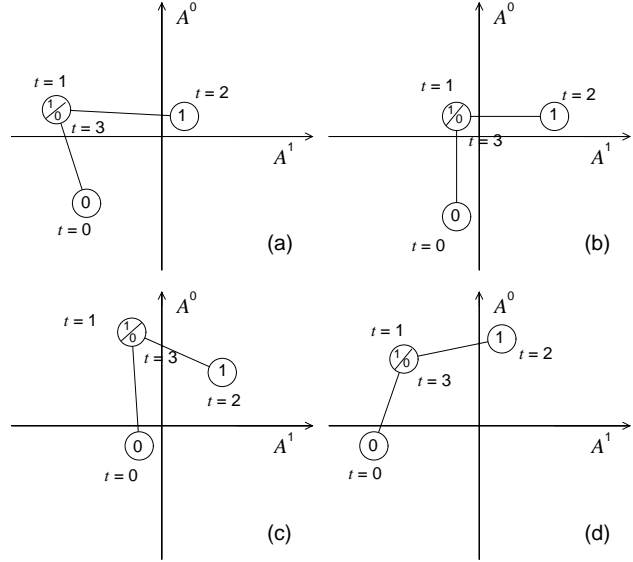,width=0.5\linewidth}}
\caption{\label{m1attfig}
(a-d) The 4 attractors for $m=1$ and $s=2$
in the multinomial regime.
The time steps are relabeled
with $t=0$ corresponding to the state with $\mu^*(t)=0$ and $\mu^*(t+1)=1$.}
\end{figure}

Consider the case $m=1$.
Eqs.~(\ref{stephist}) and (\ref{stepnonh}) show that the step size
$\langle \Delta A^\mu(t)\rangle\sim{\cal O}(1)$ and is thus self-averaging.
Since $A^\mu(0)$ is Gaussian with variance $N^{-1}$,
the values of $A^\mu(t)$ at the attractors 
can be computed to ${\cal O}(1)$.
Depending on the initial position ${\bf A}(0)\equiv(A^1(0), A^0(0))$,
4 attractors can be identified. 
For example, if ${\bf A}(0)$ lies in the first quadrant,
and the initial historical state is 0,
then the payoff components ${\bf k}\equiv(k^1(t), k^0(t))$
at the attractor are given by
${\bf k}(0)=(0, 0)$, ${\bf k}(1)=(-1, 0)$, ${\bf k}(2)=(-1, -1)$, 
${\bf k}(3)=(-1, 0)$,
provided that when $\Delta A^\mu(t)=0$ to order 1,
$\Delta A^\mu(t)$ is also equal to 0 to order $N^{-1/2}$.
Analysis can be simplified by noting that
when the payoff components $k_\mu(t)$
are restricted to the values 0 and $\pm1$,
Eq.~(\ref{stepsum}) can be written as
\begin{equation}
        A^\mu(t)=k_\mu\int_{0}^{2\pi}\frac{d\theta}{2\pi}
        (\cos\theta)^{\left[R+1+2\sum_{\nu\neq\mu}\vert k_\nu\vert\right]}
        =k_\mu c_{\left[R+1+2\sum_{\nu\neq\mu}\vert k_\nu\vert\right]},
\end{equation}
where $c_n\equiv 2^{-n}\binom{n}{n/2}$ for even integer $n$, 
and we have used the facts that $A^\mu(t)$ is self-averaging,
$A^\mu(0)\sim N^{-1/2}$. 
The locations of the 4 attractors are shown in Fig.~\ref{m1attfig} 
and summarised in Table~\ref{m1attstate}.

The variance of $A^\mu(t)$ of the historical states $\mu=\mu^*(t)$,
averaged over the period for each of the 4 attractors, 
can be obtained from Table~\ref{m1attstate}.
The variance of decisions in Eq.~(\ref{var}),
averaged over the 4 attractors, is then given by
\begin{equation}
        \frac{\sigma^2}{N}
        =\frac{N}{128}(7c_{R+1}^2-2c_{R+1}c_{R+3}+7c_{R+3}^2).
\end{equation}
The theoretical values are compared with simulation results
for the first 3 points of each curve corresponding to given values of $N$ 
in Fig.~\ref{m1diversity}.
The agreement is excellent.
Note that the variance in this regime deviates from the scaling relation 
with $\rho^{-1}$ in the next regime,
as evident from the splaying down from the linear relation 
in Fig.~\ref{m1diversity}.
However, when $R\gg1$, $c_{R+1}\approx c_{R+3}\approx \sqrt{2/\pi R}$,
$\sigma^2/N$ reduces to $3/16\pi\rho$,                   
showing that the deviation from the $\rho^{-1}$ scaling gradually vanishes.

\begin{table}[htp]
\begin{center}
\begin{tabular}{|c|cc|cc|cc|cc|} \hline

~$t$~     & ~$k_1(t)$~ & ~$k_0(t)$~
          & ~$A^1(t)$~ & ~$A^0(t)$~
          & ~$k_1(t)$~ & ~$k_0(t)$~
          & ~$A^1(t)$~ & ~$A^0(t)$~\\\hline

          & \multicolumn{4}{c|} {(a)}
          & \multicolumn{4}{c|} {(b)}\\\hline

~0~       & $-1$             & $-1$
          & $-c_{R+3}$       & ${-c_{R+3}}^*$  
          & 0                & $-1$
          & $0^\pm$          & ${-c_{R+1}}^*$\\  
~1~       & $-1$             & 0
          & ${-c_{R+1}}^*$   & $0^+$  
          & 0                & 0
          & ${0^-}^*$        & $0^+$\\ 
~2~       & 0                & 0
          & ${0^+}^*$        & $0^\pm$  
          & 1                & 0
          & ${c_{R+1}}^*$    & $0^\pm$\\  
~3~       & $-1$             & 0
          & $-c_{R+1}$       & ${0^+}^*$  
          & 0                & 0
          & $0^-$            & ${0^+}^*$\\\hline  

          & \multicolumn{4}{c|} {(c)}
          & \multicolumn{4}{c|} {(d)}\\\hline

~0~       & 0                & 0
          & $0^\pm$          & ${0^-}^*$  
          & $-1$             & 0
          & $-c_{R+1}$       & ${0^-}^*$\\  
~1~       & 0                & 1
          & ${0^-}^*$        & $c_{R+1}$  
          & $-1$             & 1
          & ${-c_{R+3}}^*$   & $c_{R+3}$\\  
~2~       & 1                & 1
          & ${c_{R+3}}^*$    & $c_{R+3}$  
          & 0                & 1
          & ${0^+}^*$        & $c_{R+1}$\\  
~3~       & 0                & 1
          & $0^-$            & ${c_{R+1}}^*$  
          & $-1$             & 1
          & $-c_{R+3}$       & ${c_{R+3}}^*$\\\hline  

\end{tabular}
\end{center}
\caption{\label{m1attstate}
The 4 attractors for $m=1$, $s=2$
in the multinomial regime.
In Tables~\ref{m1attstate} and \ref{m2attstate},
the time steps are relabeled with $t=0$
corresponding to the state with $\mu^*(t)=0$ and $\mu^*(t+1)=1$,
the superscripts $\pm$ of the value 0
indicate the possible signs to order $N^{-1/2}$,
and $A^\mu(t)$ with asterisks correspond to the historical states,
which are used to compute the variance of decisions
in Eq.~(\ref{var}).}
\end{table}

Now consider the case $m=2$.
Starting from initial positions 
near the origin of the 4-dimensional phase space,
we consider the attractors resulting from the 16 quadrants 
and 4 initial states.
We find 16 attractors for the attractor sequence in Eq.~(\ref{m2first}).
The positions of one of the attractors are summarised 
in Table~\ref{m2attstate},
and the values of $A^\mu(t)$ for the historical states $\mu=\mu^*(t)$,
which are used to compute the variance of decisions in Eq.~(\ref{var}) 
are summarised in Table~\ref{m2atthis}.
Averaging over the period and over the attractors,
the variance of decisions in Eq.~(\ref{var}) becomes
\begin{eqnarray}
        \frac{\sigma^2}{N}
        &=&\frac{N}{1024}(14c_{R+7}^2+41c_{R+5}^2+42c_{R+3}^2
        +15c_{R+1}^2\nonumber\\
        & &+2c_{R+7}c_{R+5}-2c_{R+7}c_{R+3}+2c_{R+5}c_{R+3}
        -2c_{R+5}c_{R+1}).
\end{eqnarray}
Since the attractor sequence in Eq.~(\ref{m2second}) 
is related to Eq.~(\ref{m2first}) by conjugation symmetry,
this expression is already the sample average of the variance.
Again, the theoretical values of the first 3 points of each curve 
in Fig.~\ref{m2diversity} have an excellent agreement 
with the simulation results,
and deviates from the $\rho^{-1}$ scaling in the next regime.
When $R\gg1$, $c_{R+1}\approx c_{R+3}\approx c_{R+5}\approx c_{R+7}
\approx\sqrt{2/\pi R}$,
$\sigma^2/N$ approaches $7/32\pi\rho$.

\begin{table}[htp]
\begin{center}
\begin{tabular}{|c|cccc|cccc|} \hline

~$t$~ & ~$k_0$~       & ~$k_1$~       & ~$k_2$~       & ~$k_3$~
      & ~~$A^0$~~     & ~~$A^1$~~     & ~~$A^2$~~     & ~~$A^3$~~\\\hline
0     & 0             & 0             & 0             & 0
      & ${0^-}^*$     & $0^\pm$       & $0^\pm$       & $0^\pm$\\
1     & 1             & 0             & 0             & 0
      & $c_{R+1}$     & ${0^-}^*$     & $0^\pm$       & $0^\pm$\\
2     & 1             & 1             & 0             & 0
      & $c_{R+3}$     & $c_{R+3}$     & $0^\pm$       & ${0^-}^*$\\
3     & 1             & 1             & 0             & 1
      & $c_{R+5}$     & $c_{R+5}$     & $0^\pm$       & ${c_{R+5}}^*$\\
4     & 1             & 1             & 0             & 0
      & $c_{R+3}$     & $c_{R+3}$     & ${0^-}^*$     & $0^\pm$\\
5     & 1             & 1             & 1             & 0
      & $c_{R+5}$     & ${c_{R+5}}^*$ & $c_{R+5}$     & $0^\pm$\\
6     & 1             & 0             & 1             & 0
      & $c_{R+3}$     & $0^\pm$       & ${c_{R+3}}^*$ & $0^\pm$\\
7     & 1             & 0             & 0             & 0
      & ${c_{R+1}}^*$ & $0^\pm$       & $0^\pm$       & $0^\pm$\\\hline

\end{tabular}
\end{center}
\caption{\label{m2attstate}
An atractor for $m=2$, $s=2$ in the multinomial regime
with the sequence in Eq.~(\ref{m2first}).}
\end{table}

\begin{table}[htp]
\begin{center}
\begin{tabular}{|c|cccccccc|}\hline

Attractor      & 1  & 2  & 3  & 4 
               & 5  & 6  & 7  & 8\\\hline

$\mu^*(0)=0$   & $0^-$ & $0^-$ & $0^-$ & $0^-$
               & $0^-$ & $0^-$ & $0^-$ & $0^-$\\

$\mu^*(1)=1$   & $0^-$ & $0^-$ & $0^-$ & $0^-$
               & $-c_{R+3}$ & $-c_{R+5}$ & $-c_{R+5}$ & $-c_{R+7}$\\

$\mu^*(2)=3$   & $0^-$ & $-c_{R+5}$ & $0^-$ & $-c_{R+7}$ 
               & $0^-$ & $-c_{R+3}$ & $0^-$ & $-c_{R+5}$\\

$\mu^*(3)=3$   & $c_{R+5}$ & $0^+$ & $c_{R+7}$ & $0^+$ 
               & $c_{R+3}$ & $0^+$ & $c_{R+5}$ & $0^+$\\

$\mu^*(4)=2$   & $0^-$ & $0^-$ & $-c_{R+5}$ & $-c_{R+7}$ 
               & $0^-$ & $0^-$ & $-c_{R+3}$ & $-c_{R+5}$\\

$\mu^*(5)=1$   & $c_{R+5}$ & $c_{R+7}$ & $c_{R+3}$ & $c_{R+5}$
               & $0^+$ & $0^+$ & $0^+$ & $0^+$\\

$\mu^*(6)=2$   & $c_{R+3}$ & $c_{R+5}$ & $0^+$ & $0^+$
               & $c_{R+5}$ & $c_{R+7}$ & $0^+$ & $0^+$\\

$\mu^*(7)=0$   & $c_{R+1}$ & $c_{R+3}$ & $c_{R+3}$ & $c_{R+5}$
               & $c_{R+3}$ & $c_{R+5}$ & $c_{R+5}$ & $c_{R+7}$\\\hline\hline

Attractor      & 9  & 10 & 11 & 12
               & 13 & 14 & 15 & 16\\\hline

$\mu^*(0)=0$   & $-c_{R+1}$ & $-c_{R+3}$ & $-c_{R+3}$ & $-c_{R+5}$
               & $-c_{R+3}$ & $-c_{R+5}$ & $-c_{R+5}$ & $-c_{R+7}$\\ 

$\mu^*(1)=1$   & $0^-$ & $0^-$ & $0^-$ & $0^-$
               & $-c_{R+1}$ & $-c_{R+3}$ & $-c_{R+3}$ & $-c_{R+5}$\\

$\mu^*(2)=3$   & $0^-$ & $-c_{R+3}$ & $0^-$ & $-c_{R+5}$
               & $0^-$ & $-c_{R+1}$ & $0^-$ & $-c_{R+3}$\\

$\mu^*(3)=3$   & $c_{R+3}$ & $0^+$ & $c_{R+5}$ & $0^+$ 
               & $c_{R+1}$ & $0^+$ & $c_{R+3}$ & $0^+$\\ 

$\mu^*(4)=2$   & $0^-$ & $0^-$ & $-c_{R+3}$ & $-c_{R+5}$ 
               & $0^-$ & $0^-$ & $-c_{R+1}$ & $-c_{R+3}$\\ 

$\mu^*(5)=1$   & $c_{R+3}$ & $c_{R+5}$ & $c_{R+1}$ & $c_{R+3}$
               & $0^+$ & $0^+$ & $0^+$ & $0^+$\\

$\mu^*(6)=2$   & $c_{R+1}$ & $c_{R+3}$ & $0^+$ & $0^+$
               & $c_{R+3}$ & $c_{R+5}$ & $0^+$ & $0^+$\\

$\mu^*(7)=0$   & $0^+$ & $0^+$ & $0^+$ & $0^+$
               & $0^+$ & $0^+$ & $0^+$ & $0^+$\\\hline

\end{tabular}
\end{center}
\caption{\label{m2atthis}
The values of $A^\mu(t)$ for the historical states $\mu=\mu^*(t)$
for the attractors with $m=2$, $s=2$ in the multinomial regime
in Eq.~(\ref{m2first}).
The time steps are relabeled
with $t=0$ corresponding to the state with $\mu^*(t)=0$ and $\mu^*(t+1)=1$,
the superscripts $\pm$ of the value $0$
indicate the signs to order $N^{-1/2}$.}
\end{table}

The variance of decisions for higher values of $m$ 
can be obtained by exhaustive computer search 
starting from the $2^D$ quadrants of the phase space 
and the $D$ initial states.
Since the number of cases grows rapidly with $D$,
one may use a Monte Carlo sampling of the initial conditions 
to determine the variance.

Before we close this section,
we remark that the periodic average of the decisions $A^\mu(t)$ 
at the historical states $\mu=\mu^*(t)$ have a vanishing sample average,
but the periodic average does not necessarily vanish for individual samples.
For example,
the attractor (a) in Table~\ref{m1attstate} has a periodic average of 
$\langle A^\mu(t)\rangle=-(c_{R+1}+c_{R+3})/2$
at the historical states $\mu=\mu^*(t)$.
The variance is often regarded as a measure of the system efficiency,
based on the observation that the average decisions vanish 
at high values of $m$ \cite{minority,symmetry,savit}.
However, this is not the case for the low values of $m$ we are studying.
In the context of market modeling,
a nonzero periodic average of decisions 
indicates the existence of arbitrage opportunities,
and in the context of modeling multi-agent control,
it means that there is an imbalance in the utilization of resources.
Hence the variance cannot be regarded as an intrinsic measure 
of global efficiency.
Nevertheless,
the phase space motion points in the direction of reducing the winning margin,
as seen in Eq.~(\ref{stepchange}),
which traps the attractors around the origin,
as shown in Figs.~\ref{phase} and \ref{m1attfig}.
As a result, the average of decisions is bounded 
by the step sizes at the attractor,
so that small variances also imply small averages,
and the variance can still be considered 
as a good approximate measure of efficiency.

\section{\label{sec:scaling}The Scaling Regime}

When $\rho\sim 1$, the clusters of agents making identical decisions 
effectively become continuously distributed in their preference of strategies.
Since the shift of preferences at the attractor is much narrower 
than the spread-out preference distribution,
the size of the clusters switching strategies is effectively independent 
of the detailed profile of the preference distribution.
For generic preference distributions, the width scales as $\sqrt{R}$,
and hence the size of typical clusters scales as $R^{-1/2}$.
This leads to the scaling of the variance $\sigma^2/N\sim\rho^{-1}$ 
\cite{compare2}.
Compared with the typical cluster size of scaling as $N$ 
in the multinomial regime,
the typical cluster size in the scaling regime only scales as $\sqrt{N}$.
Nevertheless, it is sufficiently numerous 
that agent cooperation in this regime can be described at the level of
statistical distributions of strategy preference,
resulting in the scaling relation.

In the integral of Eq.~(\ref{stephist}),
significant contributions only come from 
$\theta\sim1/\sqrt{R}$ or $\theta-\pi\sim1/\sqrt{R}$,
so that the factor $\cos^R\theta$
can be approximated by $\exp(-R\theta^2/2)$.
This simplifies Eq.~(\ref{stephist}) to
\begin{equation}
        \langle\Delta A^\mu(t)\rangle
        =-\sqrt{\frac{2}{\pi R}}{\rm sgn}A^\mu(t)
\label{stepav}
\end{equation}
for $\mu=\mu^*(t)$.
Since the step sizes scale as $R^{-1/2}$,
they remain self-averaging.
Similarly,
$\langle\Delta A^\mu(t)\rangle=0$ using Eq.~(\ref{stepnonh}).
The 2 cases can be summarized as
\begin{equation}\label{stepscale}
        \Delta A^\mu(t)
        =-\delta_{\mu,\mu^*(t)}\sqrt{\frac{2}{\pi R}}{\rm sgn}A^\mu(t).
\end{equation}
This result shows that the preference distribution
among agents of a given pair
is effectively a Gaussian with variance $R$,
so that the number of agents switching strategies at time $t$
scales as 2 times the height of the Gaussian distribution 
(2 being the shift of preference per step),
which is $\sqrt{2/\pi R}$.
Thus by spreading the preference distribution,
diversity reduces the step size and hence maladaptation.

As a result of Eq.~(\ref{stepscale}),
the motion in the phase space is rectilinear,
each step only making a move of fixed size 
along the direction of the historical state.
Consequently,
each state of the attractor is confined in a $D$-dimensional hypercube
of size $\sqrt{2/\pi R}$,
irrespective of the initial position of the $A^\mu$ components.
This confinement enables us to compute the variance of the decisions.
Without loss of generality,
let us relabel the time steps in the periodic attractor,
with $t=0$ corresponding to the state with $\mu^*(t)=0$ and $\mu^*(t+1)=1$.
We denote as $t_\mu$ the step at which state $\mu$ first appears
in the relabeled sequence.
(For example, $t_0=0$, $t_1=1$, $t_2=4$ and $t_3=2$
for the attractor sequence in Eq.~(\ref{m2first}).)

When state $\mu$ first appears in the attractor on or after $t=0$,
the winning state is $\sigma (t_\mu)$.
Furthermore,
since there is no phase space motion along the nonhistorical directions,
$A^\mu(t_\mu)=A^\mu(0)$.
Since the winning state is determined by the minority decision,
we have $A^\mu(0)[2\sigma(t_\mu)-1]<0$.
Similarly, when state $\mu$ appears in the attractor the second time,
the winning state is $1-\sigma (t_\mu)$,
and $A^\mu(t)=A^\mu(0)+[2\sigma (t_\mu)-1]\sqrt{2/\pi R}$.
The winning condition imposes that $A^\mu(t)[1-2\sigma(t_\mu)]<0$.
Combining,
\begin{equation}
        -\sqrt{\frac{2}{\pi R}}<A^\mu(0)[2\sigma (t_\mu)-1]<0.
\end{equation}
Suppose the game starts from the initial state $A^\mu_0$,
which are Gaussian variables with mean 0 and variance $1/N$.
They change in steps of size $\sqrt{2/\pi R}$ until they reach the attractor,
whose $2D$ historical states are then given by
\begin{equation}
        \sqrt{\frac{2}{\pi R}}
        {\rm frac}\left(\sqrt{\frac{\pi R}{2}}A^\mu_0\right)
        ~~~~{\rm and}~~~~
        \sqrt{\frac{2}{\pi R}}\left\{
	{\rm frac}\left(\sqrt{\frac{\pi R}{2}}A^\mu_0\right)-1\right\},
\end{equation}
where ${\rm frac}(x)$ represents the decimal part of $x$.
Using Eq.~(\ref{var}),
this corresponds to a variance of decisions 
given by $\sigma^2/N=f(\rho)/2\pi\rho$,
where 
\begin{eqnarray}
        f(\rho)=&&\Biggl\langle\frac{1}{D}\sum_{\mu=0}^{D-1}\left\{
        \left[{\rm frac}\left(\sqrt{\frac{\pi R}{2}}A^\mu_0\right)\right]^2
        -{\rm frac}\left(\sqrt{\frac{\pi R}{2}}A^\mu_0\right)
        +\frac{1}{2}\right\}\nonumber\\
        &&-\left\{\frac{1}{D}\sum_{\mu=0}^{D-1}
        \left[{\rm frac}\left(\sqrt{\frac{\pi R}{2}}A^\mu_0\right)
        -\frac{1}{2}\right]\right\}^2\Biggr\rangle.
\end{eqnarray}
Since $A^\mu_0$ are independent variables,
$f(\rho)$ is simplified to
\begin{equation}
        f(\rho)=\left(1-\frac{1}{D}\right)\left\langle\left[
        {\rm frac}\left(\sqrt{\frac{\pi R}{2}}A^\mu_0\right)\right]^2
	\right\rangle
        +\frac{1}{D}\left\langle{\rm frac}\left(\sqrt{\frac{\pi R}{2}}
        A^\mu_0\right)\right\rangle^2.
\end{equation}
Since $A^\mu_0$ are Gaussian variables with mean 0 and variance $N^{-1}$, 
we have
\begin{equation}
        \left\langle\left[{\rm frac}\left(\sqrt{\frac{\pi R}{2}}
        A^\mu_0\right)\right]^n\right\rangle
        =\int_{0}^{1}d\xi\left[\sum_{r=-\infty}^{\infty}
        \frac{e^{-\frac{(r+\xi)^2}{\pi\rho}}}{\sqrt{\pi^2\rho}}\right]\xi^n.
\end{equation}
When $\rho\ll 1$,
the integrals are dominated by peaks at $\xi=0$ and $\xi=1$,
yielding $\langle{\rm frac}(\sqrt{\pi R/2}A^\mu_0)\rangle
=\langle{[\rm frac}(\sqrt{\pi R/2}A^\mu_0)]^2\rangle=1/2$.
As a result, $f(\rho)=(1-1/2D)/2$.
On the other hand,
when $\rho\gg 1$,
the step sizes become much smaller than the variance of $A^\mu_0$,
so that ${\rm frac}(\sqrt{\pi R/2}A^\mu_0)$
becomes a uniform distribution between 0 and 1, leading to
$\langle{\rm frac}(\sqrt{\pi R/2}A^\mu_0)\rangle=1/2$ and
$\langle{[\rm frac}(\sqrt{\pi R/2}A^\mu_0)]^2\rangle=1/3$,
resulting in $(1-1/4D)/3$ for $\rho\gg 1$.
Hence $f(\rho)$ is a smooth function of $\rho$ varying, 
for example, from $3/8$ to $7/24$ for $m=1$.
Thus $\sigma^2/N$ depends on $\rho$
mainly through the step size factor $1/2\pi\rho$,
whereas $f(\rho)$ merely provides a higher order correction
to the functional dependence.
This accounts for the scaling regime 
in Figs.~\ref{m1diversity} and \ref{m2diversity}.  
Furthermore, we note that $f(\rho)$ rapidly approaches $1/3$ 
when $m$ increases.
Hence for general values of $D$,
$\sigma^2/N\rightarrow 1/6\pi\rho$,
provided that $m$ is not too large.
This leads to the data collapse of the variance for $m=1$ and $m=2$
in the inset of Fig.~\ref{m2diversity}.

Analogous to the multinomial regime,
the hypercube picture implies 
that both the standard deviation and the average of $A^\mu$ 
are bounded by the step size.
Hence the variance is a sufficient measure of system efficiency.

This result can be compared with that in \cite{moro},
where it was found that the variance scales as $\alpha^{1/2}$ 
in the presence of random initial conditions.
A similar $\alpha^{1/2}$ scaling was also reported for the batch MG 
\cite{heimel}.
Their results are different from ours that the variance 
is effectively independent of $D$ (where $\alpha=D/N$).
However, the simulation data in Fig.~\ref{complexity} indicates 
that the difference may not be in conflict with each other.
For a sufficiently large value of $\rho$, say $\rho=16$, 
the data in the regime immediately below $\alpha_c$ appears to be consistent
with a power-law dependence with an exponent approaching 0.5,
as predicted by \cite{moro,heimel}.
When $\alpha$ reaches lower values,
the variance flattens out,
showing that our results are applicable to the regime 
of $m$ being not too large.

\section{\label{sec:kinetic}The Kinetic Sampling Regime}

When $\rho\sim N$,
the average step sizes scale as $N^{-1}$
and are no longer self-averaging.
Rather, Eq.~(\ref{stepchange}) shows that the size of a step 
along the direction of historical states at time $t$
is $2/N$ times the number of agents who switch strategies at time $t$,
which is Poisson distributed with a mean $\Delta/2$,
implied by Eq.~(\ref{stepav}). 
Here $\Delta$ is the average step size given by Eq.~(\ref{stepdef}).
However, since the attractor is formed by steps
which {\it reverse the sign of} $A^\mu$,
the average step size in the attractor is {\it larger}
than that in the transient state,
because a long jump is the vicinity of the attractor is more likely 
to get trapped.

\begin{figure}
\centerline{\epsfig{figure=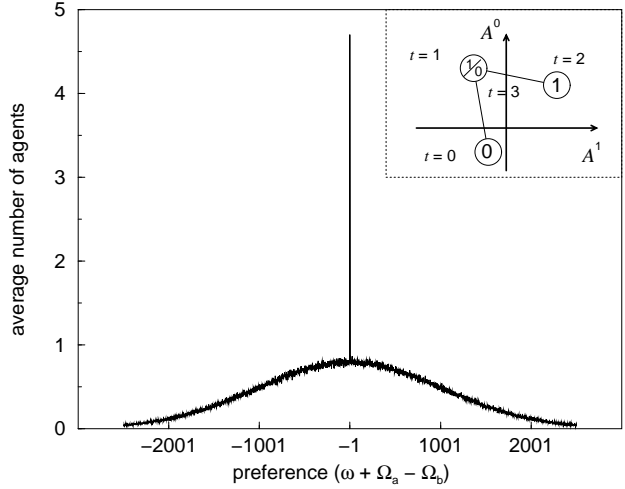,width=0.5\linewidth}}
\caption{\label{m1evidence}
Experimental evidence of the kinetic sampling effect for $m=1$:
steady-state preference distribution
of the average number of agents
holding the identity strategy and its complement,
immediately before $t=0$, and $\rho=N=1023$
and averaged over 100000 samples.
Inset: The labeling of the time steps in the attractor.}
\end{figure}

\begin{figure}
\centerline{\epsfig{figure=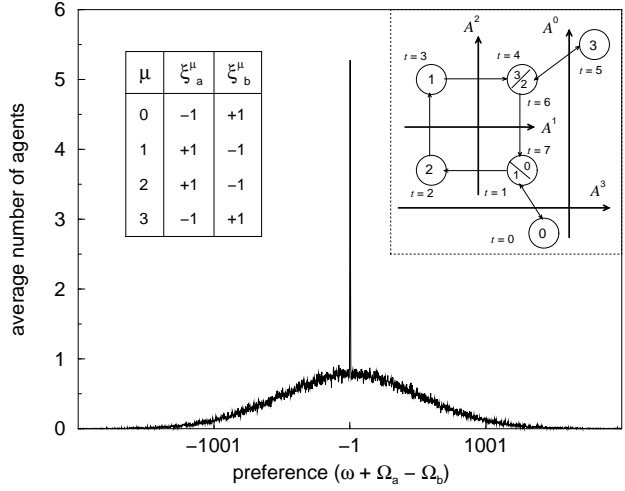,width=0.5\linewidth}}
\caption{\label{m2evidence}
Experimental evidence of the kinetic sampling effect for $m=2$: 
steady-state preference distribution
of the average number of agents
holding the XOR strategy $\xi_a^\mu$ and its complement $\xi_b^\mu$,
immediately before $t=4$, and $\rho=N=511$ 
and averaged over 50000 samples.
Inset: The labeling of the time steps in the attractor.}
\end{figure}

To consider the origin of this effect,
we focus in Fig.~\ref{m1evidence} on how the average number of agents,
who hold the identity strategy with $\sigma_a^\mu=\mu$ 
and its complementary strategy $\sigma_b^\mu=1-\mu$,
depends on the preference $\omega+\Omega_a-\Omega_b$, 
when the system reaches the steady state 
in games with $m=1$. 
Since the preferences are time dependent,
we sample their frequencies at a fixed time, say,
immediately before $t=0$ in the inset of Fig.~\ref{m1evidence}. 
One would expect that the bias distribution is reproduced.
However, we find that a sharp peak exists at $\omega+\Omega_a-\Omega_b=-1$.
This value of the preference corresponds to that of the attractor step 
from $t=3$ to $t=0$, 
when at state 0, decision 0 wins and decision 1 loses, 
and $\omega+\Omega_a-\Omega_b$ changes from $-1$ to $+1$.
The peak at the attractor step shows that its average step 
is self-organized to be larger
than those of the transient steps described by the background distribution.
Similarly for $m=2$, Fig.~\ref{m2evidence} shows 
the average number of agents who hold the XOR strategy $\xi_a^\mu$ 
and its complement $\xi_b^\mu=-\xi_a^\mu$ 
when the attractor sequence is Eq.~(\ref{m2second}). 
At the attractor step immediately before $t=4$ 
in the inset of Fig.~\ref{m2evidence}, 
the state is 1. 
Decision 1 wins and decision 0 loses, 
changing the preference $\omega+\Omega_a-\Omega_b$ from $-1$ to $+1$, 
and hence contributing to the sharp peak at $\omega+\Omega_a-\Omega_b=-1$.

This effect that favors the cooperation of larger clusters of agents
is referred to as the {\it kinetic sampling} effect.
To describe this effect,
we consider the probability of $P_{\rm att}(\Delta{\bf A})$
of step sizes $\Delta{\bf A}$ in the attractor.
For convenience,
we only consider $\Delta A^\mu>0$ for all $\mu$.
Assuming that all states of the phase space are equally likely
to be accessed by the initial condition, we have
\begin{equation}
        P_{\rm att}(\Delta{\bf A})
        =\sum_{\bf A}P_{\rm att}(\Delta{\bf A},{\bf A}),
\end{equation}
where $P_{\rm att}(\Delta{\bf A},{\bf A})$
is the probability of finding the position ${\bf A}$
with displacement $\Delta{\bf A}$ in the attractor.  
Consider the example of $m=1$,
where there is only one step along each axis $A^\mu$.
The sign reversal condition implies that
\begin{equation}
        P_{\rm att}(\Delta{\bf A},{\bf A})
        =P_{\rm Poi}(\Delta{\bf A})
        \prod_\mu\Theta[-A^\mu(A^\mu+\Delta A^\mu)],
\end{equation}
where $P_{\rm Poi}(\Delta{\bf A})$
is the Poisson distribution of step sizes, yielding
\begin{equation}
        P_{\rm att}(\Delta{\bf A})
        \propto P_{\rm Poi}(\Delta{\bf A})\prod_\mu\Delta A^\mu.
\end{equation}
We note that the extra factors of $\Delta A^\mu$ favor large step sizes.
Thus the attractor averages $\langle(\Delta A^\pm)^2\rangle_{\rm att}$,
which are required for computing the variance of decisions, are given by
\begin{equation}
        \langle(\Delta A^\pm)^2\rangle_{\rm att}=
        \frac{\langle(\Delta A^\pm)^2\Delta A^+\Delta A^-\rangle_{\rm Poi}}
        {\langle\Delta A^+\Delta A^-\rangle_{\rm Poi}}.
\label{attav}
\end{equation}
Furthermore, correlation effects come into action
when the step sizes become non-self-averaging.
There are agents who contribute to both $\Delta A^+$ and $\Delta A^-$, 
giving rise to their correlations. 
Thus, the variance of decisions is higher 
when correlation effects are considered. 
In Eq.~(\ref{stepchange}), 
the strategies of the agents contributing to $\Delta A^+$ and $\Delta A^-$ 
satisfy $\xi^+_a-\xi^+_b=\pm 2$ and $\xi^-_a-\xi^-_b=\mp 2$ respectively. 
Among the agents contributing to $\Delta A^+$, 
the extra requirement of $\xi^-_a-\xi^-_b=\mp 2$ 
implies that an average of $1/4$ of them 
also contribute to $\Delta A^-$. 
Hence, the number of agents contributing to both steps 
is a Poisson variable with mean $\Delta/8$. 
Similarly, the number of agents exclusive to the individual steps 
are Poisson variables with means $3\Delta/8$. 
Algebraically, Eq.~(\ref{stepchange}) can be decomposed as
\begin{eqnarray}
        & &\Delta A^\pm=
        \frac{2}{N}\sum_{a<b}\sum_{r=\pm1}S_{ab}(-r-\Omega_a+\Omega_b)
        \delta(\xi_a^\pm-\xi_b^\pm+2r)\delta(\xi_a^\mp-\xi_b^\mp-2r)
        \nonumber\\
        & &+\frac{2}{N}\sum_{a<b}\sum_{r=\pm1}S_{ab}(-r-\Omega_a+\Omega_b)
	\delta(\xi_a^\pm-\xi_b^\pm+2r)[\delta(\xi_a^\mp-\xi_b^\mp)
        +\delta(\xi_a^\mp-\xi_b^\mp+2r)]\}.
\label{m1parts}
\end{eqnarray}
Respectively, the first and terms are equal to 
$2/N$ times the number of agents, 
common to both steps $\Delta A^\pm$ and exclusive to the individual steps, 
with means $\Delta/8$ and $3\Delta/8$, 
as can be verified by a derivation 
similar to that of Eq.~(\ref{stephist}) from Eq.~(\ref{stepchange}).
Hence the denominator of Eq.~(\ref{attav}) is given by
\begin{equation}
        \langle\Delta A^+\Delta A^-\rangle_{\rm Poi}
        =\frac{4}{N^2}\sum_{a_0,a_+,a_-}
        \frac{e^{-\frac{\Delta}{8}}}{a_0!}
        \left(\frac{\Delta}{8}\right)^{a_0}
        \frac{e^{-\frac{3\Delta}{8}}}{a_+!}
        \left(\frac{3\Delta}{8}\right)^{a_+}
        \frac{e^{-\frac{3\Delta}{8}}}{a_-!}
        \left(\frac{3\Delta}{8}\right)^{a_-}
        (a_0+a_+)(a_0+a_-).
\label{poi}
\end{equation}
Expressing the moments of Poisson variables in terms of their means,
we arrive at
\begin{equation}
        \langle\Delta A^+\Delta A^-\rangle_{\rm Poi}
        =\frac{4}{N^2}\left[16\left(\frac{\Delta}{8}\right)^2
        +\frac{\Delta}{8}\right].
\end{equation}
Similarly, the numerator of Eq.~(\ref{attav}) is given by
\begin{equation}
        \langle(\Delta A^\pm)^2\Delta A^+\Delta A^-\rangle_{\rm Poi}
        =\frac{16}{N^4}\left[256\left(\frac{\Delta}{8}\right)^4
        +240\left(\frac{\Delta}{8}\right)^3
        +40\left(\frac{\Delta}{8}\right)^2
        +\frac{\Delta}{8}\right].
\end{equation}
Together we obtain
\begin{equation}
        \langle(\Delta A^\pm)^2\rangle_{\rm att}
        =\frac{2\Delta^3+15\Delta^2+20\Delta+4}{N^2(2\Delta+1)}.
\label{attdelta}
\end{equation}
The possible attractor states 
are given by $A^\mu=m_\mu/N$ and $m_\mu/N-\Delta A^\mu$,
where $m_\mu=1,~3,~\ldots,~N\Delta A^\mu-1$.
This yields a variance of
\begin{equation}
        \frac{\sigma^2}{N}
        =\frac{N}{4}
        \left\langle\left\{\frac{1}{D}\sum_{\mu=0}^{D-1}
        \left[\left(\frac{m_\mu}{N}\right)^2
        -\Delta A^\mu\left(\frac{m_\mu}{N}\right)
        +\frac{1}{2}(\Delta A^\mu)^2\right]
        -\left[\frac{1}{D}\sum_{\mu=0}^{D-1}
        \left(\frac{m_\mu}{N}-\frac{1}{2}\Delta A^\mu\right)
        \right]^2\right\}\right\rangle.
\end{equation}
Averaging over the attractor states, we find
\begin{equation}
        \frac{\sigma^2}{N}
        =\frac{7\langle(N\Delta A^+)^2\rangle_{\rm att}
        +7\langle(N\Delta A^-)^2\rangle_{\rm att}-8}{192N},
\label{attvar}
\end{equation}
which gives, on combining with Eq.~(\ref{attdelta}),
\begin{equation}
        \frac{\sigma^2}{N}=
        \frac{14\Delta^3+105\Delta^2+132\Delta+24}{96N(2\Delta+1)}.
\label{m1kinetic}
\end{equation}
When the diversity is low,
$\Delta\gg1$, and Eq.~(\ref{m1kinetic})
reduces to $\sigma^2/N\equiv7/48\pi\rho$,
agreeing with the scaling result of the previous section.
When $\rho\sim N$,
Eq.~(\ref{m1kinetic}) has excellent agreement with simulation results,
which significantly deviate above the scaling relation, 
as shown in Fig.~\ref{m1diversity}.

When $\rho\gg N$,
Eq.~(\ref{m1kinetic}) predicts that $\sigma^2/N$ should approach $1/4N$.
This can be explained as follows.
Analysis shows that only those agents holding 
the identity strategy and its complement can complete
both hops along the $A^\pm$ axes after they have adjusted their
preferences to $\omega+\Omega_a-\Omega_b=\pm1$.
Since there are fewer and fewer fickle agents in the limit $\rho\gg N$,
one would expect that a single agent of this type 
would dominate the game dynamics,
and $\sigma^2/N$ would approach $1/4N$.

However, as shown in Fig.~\ref{m1diversity},
the simulation data approaches the limit $0.43/N$ when $\rho\gg N$,
significantly higher than $0.25/N$.
This discrepancy requires the consideration of the {\it waiting} effect,
which has been sketched in \cite{wong},
and will be explained in details elsewhere.

\begin{figure}
\centerline{\epsfig{figure=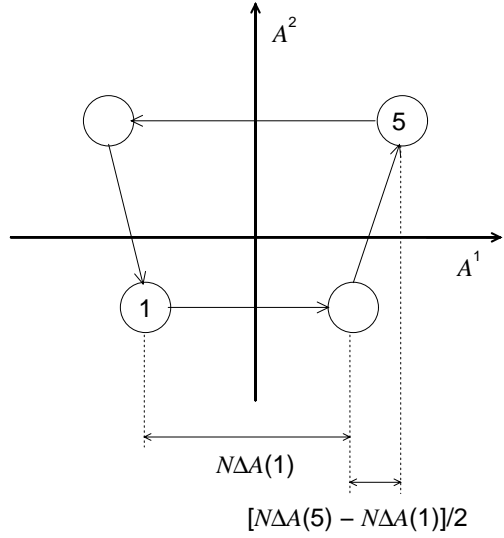,width=0.4\linewidth}}
\caption{\label{m2subspace}
The relative positions of the steps $\Delta A(1)$ and $\Delta A(5)$
for the case $\Delta A(5)>\Delta A(1)$.
Here they are shown symmetrically positioned.}
\end{figure}

Next, we turn to the kinetic sampling effects for $m=2$.
As shown in Fig.~\ref{phase}(b),
the situation is more complicated than that of $m=1$ 
since there are two steps moving along the direction $A^1$ and $A^2$.
Consider the attractor sequence in Eq.~(\ref{m2first}).
The step $\Delta A(1)$ can initiate from $A^1=m_1/N$, 
with $m_1=-1,~\ldots,~-N\Delta A(1)+1$,
where for convenience the state labels of the step sizes at time $t$
are implicitly taken to be the historical states $\mu^*(t)$.
Similarly,
the step $\Delta A(5)$ can initiate from $A^1=m_5/N$, 
with $m_5=1,~\ldots,~N\Delta A(5)-1$.
However, since the two steps are linked by steps along the direction $A^2$,
their positions are no longer independent.
Taking into consideration the many possibilities 
of their relative displacements make the problem intractable.
As shown in Fig.~\ref{m2subspace},
we only consider the most probable case 
that the two steps are symmetrically positioned,
that is, their midpoints have the same $A^1$ coordinate.
In this case, the possible initial positions of the steps are $A(1)=m_1/N$,
with $m_1=-1,~\ldots,~-[N\Delta A(1)+N\Delta A(5)]/2+1$,
and $A(5)=m_5/N$, with $m_5=m_1+[N\Delta A(1)+N\Delta A(5)]/2$.
Thus, the number of possible states 
along the direction $A^1$ is $[N\Delta A(1)+N\Delta A(5)]/4$.
Considering the motion in the 4 directions,
the total number of possible states is 
$[N\Delta A(0)/2][(N\Delta A(1)+N\Delta A(5))/4][N\Delta A(2)/2]
[(N\Delta A(4)+N\Delta A(6))/4]$.

Extending the derivation of Eq.~(\ref{attvar}) to the case of $m=2$,
we have
\begin{eqnarray}
        \frac{\sigma^2}{N}
        =\frac{N}{256}
        &&\Biggl\{5\langle\Delta A(0)^2\rangle_{\rm att}
        +5\left\langle\frac{(\Delta A(1)+\Delta A(5))^2}{4}
	\right\rangle_{\rm att}\nonumber\\
        &&+5\langle\Delta A(2)^2\rangle_{\rm att}
        +5\left\langle\frac{(\Delta A(4)+\Delta A(6))^2}{4}
	\right\rangle_{\rm att}-16\Biggr\},
\label{m2var}
\end{eqnarray}
where the attractor averages are defined as the Poisson averages 
weighted by kinetic sampling.
For example,
\begin{equation}
        \langle\Delta A(0)^2\rangle_{\rm att}
        =\frac{\langle\Delta A(0)[\Delta A(1)+\Delta A(5)]
        \Delta A(2)[\Delta A(4)+\Delta A(6)]\Delta A(0)^2\rangle_{\rm Poi}}
        {\langle\Delta A(0)[\Delta A(1)+\Delta A(5)]
        \Delta A(2)[\Delta A(4)+\Delta A(6)]\rangle_{\rm Poi}}.
\end{equation}
This requires us to compute Poisson averages such as
$\langle\Delta A(t_1)\cdots\Delta A(t_k)\rangle_{\rm Poi}$.
The following identity for Poisson averages is useful.
Consider a universal set of $M$ elements,
and the sizes of the sets $B_1\cdots B_k$ and their intersections 
are Poisson distributed.
Then the expectation of the product 
$\vert B_1\vert\cdots\vert B_k\vert$ is given by
\begin{equation}
        \langle\vert B_1\vert\cdots\vert B_k\vert\rangle
        =\prod_{r=1}^k\langle\vert B_r\vert\rangle
        +\sum_{r<s}\langle\vert B_r\cap B_s\vert\rangle
        \prod_{u\neq rs}\langle\vert B_u\vert\rangle+\cdots
        +\langle\vert\bigcap_{r=1}^k B_r\vert\rangle.
\label{iden}
\end{equation}
This identity can be proved by writing
\begin{equation}
        \vert B_1\vert\cdots\vert B_k\vert
        =\sum_{i_1=1}^M\cdots\sum_{i_k=1}^M
        \Theta(i_1\in B_1)\cdots\Theta(i_k\in B_k)
\end{equation}
where $\Theta(i_r\in B_r)$ if $i_r\in B_r$ and 0 otherwise.
In the limit of $M$ approaching infinity,
the case that all $i_r$ are distinct 
yields the expectation value in the first term of Eq.~(\ref{iden}),
the case that $i_r=i_s$ corresponds to the second term,
and the case that all $i_r$ are identical corresponds to the last term,
and so on.

Therefore, we can write
\begin{equation}
        \langle\Delta A(1)\cdots\Delta A(k)\rangle
        =\left(\frac{2}{N}\right)^k\left\{\prod_{r=1}^k b_r
        +\sum_{r<s}b_{rs}\prod_{u\neq rs}b_a+\cdots+b_{1\cdots+k}\right\}
\end{equation}
where $b_{r_1\cdots r_i}$ is the average number of agents 
simultaneously contributing to the steps
$\Delta A(r_1)\cdots\Delta A(r_i)$.

Consider the attractor sequence in Eq.~(\ref{m2first}).
Tracing the time evolution of the cumulative payoffs,
the step sizes at $t=2$ and $t=6$, for example,
are given by
\begin{equation}
        \Delta A(2)=\frac{2}{N}\sum_{a<b}\sum_{r=\pm1}S_{ab}(-r-\Omega_a(2)
        +\Omega_b(2))\delta(\xi_a^3-\xi_b^3-2r),
\end{equation}
\begin{equation}
        \Delta A(6)=\frac{2}{N}\sum_{a<b}\sum_{r'=\pm1}
        S_{ab}(-r'-\Omega_a(2)+\Omega_b(2)
        +\xi_a^1-\xi_b^1-\xi_a^2+\xi_b^2))
        \delta(\xi_a^2-\xi_a^2+2r').
\end{equation}
Following the analysis of Eq.~(\ref{m1parts}),
we find $b_2=b_6=\Delta/2$.
To find $b_{26}$,
we note that the agents shared by the two steps satisfy 
either $r=r'$ and $\xi_a^1-\xi_b^1=\xi_a^2-\xi_b^2=-2r$,
or $r=-r'$ and $\xi_a^1-\xi_b^1=0$, $\xi_a^2-\xi_b^2=2r$.
This leads to
\begin{eqnarray}
        b_{26}=
        \sum_{a<b}\sum_{r=\pm1}&&\langle
        S_{ab}(-r-\Omega_a(2)+\Omega_b(2))\rangle
        \delta(\xi_a^3-\xi_b^3-2r)\nonumber \\
        &&\times
        \{\delta(\xi_a^1-\xi_b^1+2r)\delta(\xi_a^2-\xi_b^2+2r)
        +\delta(\xi_a^1-\xi_b^1)\delta(\xi_a^2-\xi_b^2-2r)\}.
\end{eqnarray}
The two terms in this expression
consist of the contributions to $\Delta A(2)$,
with the extra restrictions of $\xi_a^1-\xi_b^1=\xi_a^2-\xi_b^2=-2r$,
or $\xi_a^1-\xi_b^1=0$ and $\xi_a^2-\xi_b^2=2r$ respectively.
Since $\xi_a^\mu-\xi_b^\mu=\pm2r$ and 0 
with probabilities $1/4$ and $1/2$ respectively,
we get $b_{26}=3\Delta/32$.
Other parameters are listed in Table~\ref{m2sets}.
This enables us to find
\begin{eqnarray}
        & &\langle\Delta A(0)[\Delta A(1)+\Delta A(5)]
        [\Delta A(4)+\Delta A(6)]\Delta A(2)\rangle_{\rm Poi}\nonumber\\
        &=&\frac{1}{8N^4}\left(32\Delta^4+84\Delta^3+\frac{169}{4}\Delta^2
        +2\Delta\right).
\end{eqnarray}
Other expressions appearing in Eq.~(\ref{m2var}) can be found similarly.
The final result is
\begin{equation}
        \frac{\sigma^2}{N}
        =\frac{160\Delta^5+1680\Delta^4+4772\Delta^3
        +\frac{272061}{64}\Delta^2+\frac{7583}{8}\Delta+17}
        {64N(32\Delta^3+84\Delta^2+\frac{169}{4}\Delta+2)}.
\label{m2kinetic}
\end{equation}
Since the attractor sequence in Eq.~(\ref{m2second}) yields the same result,
Eq.~(\ref{m2kinetic}) is the sample average of the variance.
When the diversity is low,
$\Delta\gg1$, and Eq.~(\ref{m2kinetic}) reduces to $\sigma^2/N=5/32\pi\rho$,
agreeing with the scaling result of the previous section.
When $\rho\sim N$,
Eq.~(\ref{m2kinetic})
shows that the introduction of kinetic sampling 
significantly improves the theoretical agreement with simulation results, 
as shown in Fig.~\ref{m2diversity}.
When $\rho\gg N$, Eq.~(\ref{m2kinetic}) implies that $\sigma^2/N$ 
approaches $17/128N$.
This result is not valid
since it is below the lowest possible result of $1/4N$ 
when each step is excuted by the strategy switching of only one agent.
The discrepency can be traced to the approximation 
that the average number of states along the direction $A^1$ is
$[N\Delta A(1)+N\Delta (5)]/2$, which is not precise for small steps.
For example, it can take half integer values.
We will not pursue this issue further since, in any case,
waiting effects have to be taken into account 
in analysing the case $\rho\gg N$.

\begin{table}[htp]
\begin{center}
\begin{tabular}{|c|l|} \hline

$\Delta/2$    & $b_0, b_1, b_2, b_4, b_5, b_6$\\
$\Delta/4$    & $b_{15}, b_{46}$\\
$\Delta/8$    & $b_{01}, b_{06}, b_{12}, b_{14}, b_{16}, 
		b_{24}, b_{45}, b_{56}$\\
$3\Delta/32$  & $b_{02}, b_{04}, b_{05}, b_{25}, b_{26}$\\
$\Delta/16$   & $b_{015}, b_{046}, b_{125}, b_{246}$\\
$\Delta/32$   & $b_{012}, b_{014}, b_{016}, b_{056}, 
		b_{124}, b_{126}, b_{245}$\\
$3\Delta/128$ & $b_{024}, b_{026}$\\
$\Delta/64$   & $b_{025}$\\
$\Delta/128$  & $b_{0124}, b_{0126}$\\
$\Delta/64$   & $b_{0125}, b_{0246}$\\\hline

\end{tabular}
\end{center}
\caption{\label{m2sets}
Values of $b_{t_1\cdots t_r}$
for the attractor sequence in Eq.~(\ref{m2first}).
The steps at $t=3$ and $t=4$ are identical,
so are the steps at $t=6$ and $t=7$.
Other unlisted parameters are zero.}
\end{table}

In summary, we have explained the reduction of variance
by the reduction of the fraction of fickle agents when diversity increases.
The theoretical analysis from Sections~\ref{sec:multi} to \ref{sec:kinetic}
spans the 3 regimes of small $R$,
$\rho^{-1}$ scaling, and kinetic sampling,
yielding excellent argreement with simulations over 7 decades.

It is natural to consider whether the results presented here
can be generalized to the case of the {\it exogenous} MG,
in which the information $\mu(t)$ was randomly
and independently drawn at each time step $t$ from a distribution
$\rho^\mu=1/D$ \cite{challet}.
This is different from the present {\it endogenous} version of the MG,
in which the information is determined by the sequence of the winning bits
in the game history.
The similarities and differences between the behavior of those two versions 
have been a topic of interest in the literature 
\cite{cavagna98,johnson99,challet,zheng,lee,metzler,web}.
Here we compare their behavior in games of small $m$ using the phase space 
we introduced.

\begin{figure}
\centerline{\epsfig{figure=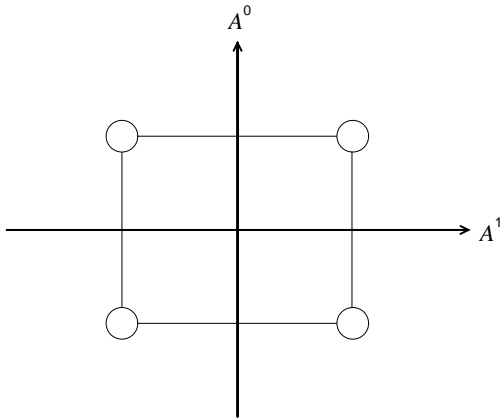,width=0.4\linewidth}}
\caption{\label{exo}
An attractor of the exogenous Minority Game for $m=1$.}
\end{figure}

In the scaling regime,
the picture that the states of the game are hopping between hypercubes
in the phase space remains valid,
as shown in Fig.~\ref{exo} for $m=1$.
At the steady state,
the attractor consists of hoppings along all edges of a hypercube,
in contrast to the endogenous case,
in which only a fraction of hypercube vertices belong to the attractor.
The behavior in the scaling regime depends 
on the scaling of the step sizes with diversity,
rather than the actual sequence of the steps.
Consequently,
the behavior is the same as the endogenous game.
In the kinetic sampling regime,
the physical picture that larger steps are more likely to be trapped 
remains valid, 
and the behavior remains qualitatively similar to the endogenous case.

\section{\label{sec:fickle}The Fraction of Fickle Agents}

\begin{figure}
\centerline{\epsfig{figure=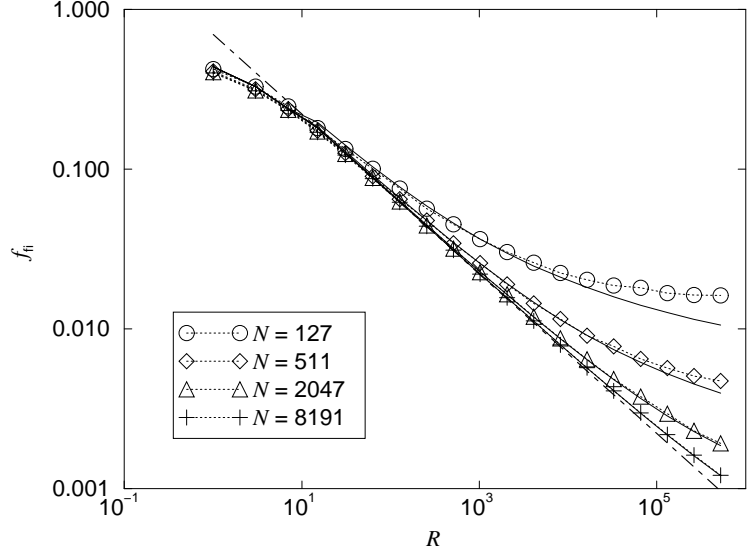,width=0.6\linewidth}}
\caption{\label{m1ficklefig}
The dependence of the fraction of fickle agents
on the randomness $R$ at $m=1$ and $s=2$.
Notations are the same as those of Fig.~\ref{m1diversity}.}
\end{figure}

\begin{figure}
\centerline{\epsfig{figure=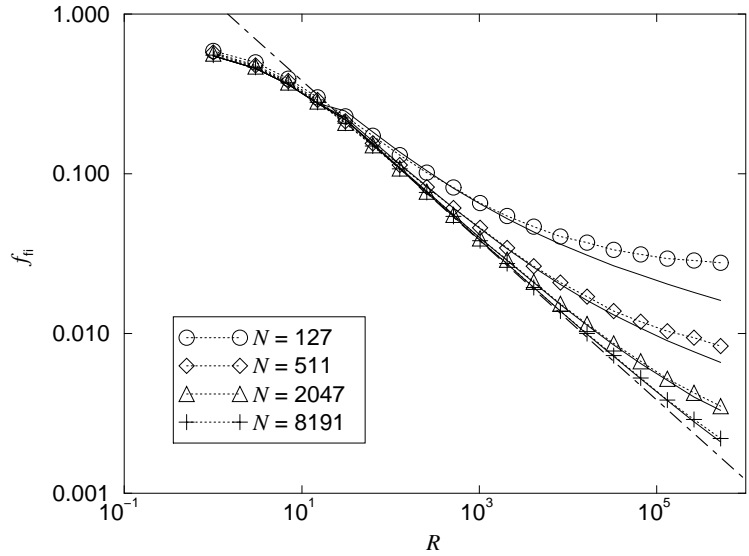,width=0.6\linewidth}}
\caption{\label{m2ficklefig}
The dependence of the fraction of fickle agents
on the randomness $R$ at $m=2$ and $s=2$.
Notations are the same as those of Fig.~\ref{m1diversity}.}
\end{figure}

This physical picture of the diversity effects is further illustrated 
by considering the fraction $f_{\rm fi}$ of fickle agents 
when the game has reached the steady state.
They hold strategy pairs whose preferences are distributed near zero,
and change sign during the attractor dynamics.
As confirmed in Figs.~\ref{m1ficklefig} and \ref{m2ficklefig},
three regimes of behavior exist.

In the multinomial regime,
we can make use of the explicit knowledge about the attractor sequence
and the evolution of the payoffs in the attractor dynamics.
Consider the example of $m=1$.
We count the type of fickle agents labeled by the strategy pairs $a<b$
and bias $\omega$ for all $t$,
with preferences
\begin{equation}
        \omega+\Omega_a(t)-\Omega_b(t)=\pm 1
        ~~~~{\rm and}~~~~
        \xi_a^\mu-\xi_b^\mu=\mp2{\rm sgn}A^\mu(t),
\end{equation}
where $\mu=\mu^*(t)$. Equivalently, we have 
\begin{equation}
        \omega=-\Omega_a(t)+\Omega_b(t)
        -\frac{1}{2}(2\sigma(t)-1)(\xi_a^{\mu^*(t)}-\xi_b^{\mu^*(t)}),
\end{equation}
where $\Omega_a(t)$ is updated by
\begin{equation}
        \Omega_a(t+1)=\Omega_a(t)+\xi_a^{\mu^*(t)}[2\sigma(t)-1].
\end{equation}
This enables us to count the types 
directly from the knowledge of the attractor sequences, 
such as Eqs.~(\ref{m1}) and (\ref{m2first}), 
without having to know the step sizes. 
Results for $m=1$ and $m=2$ are listed 
in Tables \ref{m1fickletype} and \ref{m2fickletype} respectively. 
Note that the values in the tables 
depend on the convention of ordering the strategies $a<b$, 
and here the convention of Eq.~(\ref{stlabel}) is adopted. 
Other conventions may classify the types with bias $\omega$ as $-\omega$, 
or vice versa. 
Since the average number of fickle agents of each type 
is given by Eq.~(\ref{average}),
$f_{\rm fi}$ can then be obtained by summing up the contribution 
from each type.

\begin{table}[htp]
\begin{center}
\begin{tabular}{|c|cccc|c|} \hline

$\omega$ & (a) & (b) & (c) & (d) & Total\\\hline
$-3$     &  1  &  0  &  0  &  0  & 1 \\
$-1$     &  5  &  4  &  0  &  3  & 12\\
1        &  1  &  3  &  6  &  3  & 13\\
3        &  0  &  0  &  1  &  1  & 2 \\\hline
Total    &  7  &  7  &  7  &  7  &   \\\hline

\end{tabular}
\end{center}
\caption{\label{m1fickletype}
The number of types of fickle agents
for the attractors (a)-(d) in Fig.~\ref{m1attfig}.}
\end{table}

Consider the example of $m=1$.
Table~\ref{m1fickletype} shows that there are 7 types of fickle agents 
for each attractor shown in Fig.~\ref{m1attfig}.
Averaging over initial states,
we find that an average of $25/4$ types consist of agents
with biases $\omega=\pm 1$,
and an average of $3/4$ types with $\omega=\pm 3$, 
this result being independent of the ordering of $a<b$. 
Since the average number of agents holding strategy pair $a<b$ is $N/8$, 
we have
\begin{equation}
        f_{\rm fi}=\frac{25}{32}\binom{R}{\frac{R-1}{2}}\frac{1}{2^R}
        +\frac{3}{32}\binom{R}{\frac{R-3}{2}}\frac{1}{2^R}.
\label{m1fickle}
\end{equation}
For $m=2$,
the number of types of fickle agents for the 16 attractors 
in Table~\ref{m2atthis} are listed in Table \ref{m2fickletype}.
There are 194 types of fickle agents for each attractor.
The fraction of fickle agents is given by
\begin{equation}
        f_{\rm fi}=\frac{1121}{1024}\binom{R}{\frac{R-1}{2}}\frac{1}{2^R}
        +\frac{373}{1024}\binom{R}{\frac{R-3}{2}}\frac{1}{2^R}
        +\frac{55}{1024}\binom{R}{\frac{R-5}{2}}\frac{1}{2^R}
        +\frac{3}{1024}\binom{R}{\frac{R-7}{2}}\frac{1}{2^R}.
\label{m2fickle}
\end{equation}
In the scaling regime $\rho\sim1$,
we consider the limit of $R\sim N$ in Eq.~(\ref{m1fickle}),
and obtain for $m=1$, 
\begin{equation}\label{m1 scale}
        f_{\rm fi}=\frac{7}{8}\sqrt{\frac{2}{\pi R}}.
\end{equation}
Similarly, from Eq.~(\ref{m2fickle}),
we have for $m=2$, 
\begin{equation} 
        f_{\rm fi}=\frac{97}{64}\sqrt{\frac{2}{\pi R}}.
\label{m2scale}
\end{equation}

\begin{table}[htp]
\begin{center}
\begin{tabular}{|c|rrrrrrrrrrrrrrrr|r|} \hline

Attractor & 1 & 2 & 3 & 4
          & 5 & 6 & 7 & 8
          & 9 & 10 & 11 & 12
          & 13 & 14 & 15 & 16 & Total\\\hline

$\omega=-7$   & 0 & 0 & 0 & 0
              & 0 & 0 & 0 & 0 
              & 0 & 0 & 0 & 0 
              & 0 & 0 & 0 & 1 & 1\\

$\omega=-5$   & 0 & 0 & 0 & 0 
              & 0 & 2 & 0 & 3
              & 0 & 1 & 1 & 4
              & 0 & 6 & 4 & 9 & 30\\

$\omega=-3$   & 0 & 3 & 5 & 10
              & 8 & 16 & 16 & 23
              & 7 & 20 & 22 & 28
              & 24 & 33 & 33 & 38 & 286\\
 
$\omega=-1$   & 19 & 42 & 42 & 54
              & 52 & 59 & 69 & 66
              & 76 & 73 & 76 & 75
              & 91 & 84 & 94 & 85 & 1057\\

$\omega=1$    & 120 & 87 & 92 & 71
              & 93 & 70 & 66 & 59
              & 90 & 72 & 72 & 60
              & 75 & 55 & 54 & 49 & 1185\\

$\omega=3$    & 48 & 50 & 44 & 46
              & 37 & 37 & 36 & 33           
              & 21 & 25 & 20 & 24
              & 4 & 15 & 9 & 11 & 460\\

$\omega=5$    & 7 & 11 & 10 & 12 
              & 4 & 9 & 7 & 9
              & 0 & 3 & 3 & 3 
              & 0 & 1 & 0 & 1 & 80\\

$\omega=7$    & 0 & 1 & 1 & 1
              & 0 & 1 & 0 & 1
              & 0 & 0 & 0 & 0 
              & 0 & 0 & 0 & 0 & 5\\\hline

Total         & 194 & 194 & 194 & 194
              & 194 & 194 & 194 & 194
              & 194 & 194 & 194 & 194
              & 194 & 194 & 194 & 194 & \\\hline
               
\end{tabular}
\end{center}
\caption{\label{m2fickletype}
The number of types of fickle agents
for the 16 attractors in Table~\ref{m2atthis} at $m=2$.}
\end{table}

In the kinetic sampling regime,
the fraction of fickle agents for $m=1$ 
is obtained by replacing $(\Delta A^\pm)^2$
in the numerator of Eq.~(\ref{attav}) by $(a_0+a_++a_-)/N$,
following the notation used in Eq.~(\ref{poi}). The result is
\begin{equation} 
        f_{\rm fi}=\frac{14\Delta^2+39\Delta+8}{8N(2\Delta+1)}.
\label{m1ficklek}
\end{equation}
In the limit of low diversity,
$\Delta\gg 1$ and Eq.~(\ref{m1ficklek}) reduces to Eq.~(\ref{m1 scale}).
In the limit of high diversity,
$\Delta\ll 1$ and $f_{\rm fi}$ approaches $1/N$,
implying that a single agent would dominate the game dynamics.
However,
since waiting effects are neglected,
this result is considerably lower than the simulation results.

For $m=2$, the fraction of fickle agents 
is given by the size of the union set of fickle agents at all steps,
\begin{equation}
        f_{\rm fi}=\frac{1}{N}\left\langle\sum_rb_r-\sum_{r<s}b_{rs}
        +\sum_{r<s<u}b_{rsu}\cdots\right\rangle_{\rm att}
\end{equation}
where
\begin{equation}
        \langle b_{r_1\cdots r_i}\rangle_{\rm att}
        =\frac{\langle\Delta A(0)[\Delta A(1)+\Delta A(5)]
        \Delta A(2)[\Delta A(4)+\Delta A(6)]
        b_{r_1\cdots r_i}\rangle_{\rm Poi}}
        {\langle\Delta A(0)[\Delta A(1)+\Delta A(5)]
        \Delta A(2)[\Delta A(4)+\Delta A(6)]\rangle_{\rm Poi}}.
\end{equation}
The result is
\begin{equation} 
        f_{\rm fi}=\frac{1552\Delta^4+8170\Delta^3
        +\frac{80905}{8}\Delta^2+2801\Delta+64}
        {32N(32\Delta^3+84\Delta^2+\frac{169}{4}\Delta+2)}.
\label{m2ficklek}
\end{equation}
In the limit of low diversity,
$\Delta\gg1$ and Eq.~(\ref{m2ficklek}) reduces to Eq.~(\ref{m2scale}).
In the limit of high diversity,
$f_{\rm fi}$ approaches $1/N$.
However, by tracing the types of fickle agents switching strategies 
at each time step,
one cannot find {\it any} single type of agents 
who can contribute to the dynamics of {\it all} steps.
In fact, the minimum number of agents that can complement each other 
to complete the dynamics is 2.
For example,
one agent can complete the steps at $t=0$, 1, 2, 3, 4, 
while the other one can complete the steps $t=5$, 6, 7.
Hence the asymptotic limit of $f_{\rm fi}=1/N$ is not valid.
The source of the discrepancy is the same as that for 
the invalid result of the asymptotic variance of decisions 
explained in the previous section.

As shown in Figs.~\ref{m1ficklefig} and \ref{m2ficklefig},
the theoretical predictions are confirmed by simulations,
except in the regime of extremely high diversity,
where waiting effects have to be taken into account \cite{wong}.

\section{\label{sec:convergence}Convergence Time}

Many properties of the system dependent on the transient dynamics 
also depend on its diversity.
For example, since diversity reduces the fraction of agents 
switching strategies at each time step,
it also slows down the convergence to the steady state.
Hence the convergence time increases with diversity.

We consider the example of $m=1$.
The dynamics of the game proceeds in the direction 
which reduces the variance \cite{challet}.
In the multinomial regime,
the initial position of $A^\mu$ in the phase space lies in the attractor.
Convergence to the steady state is almost instant.
Starting from the initial state 0,
the convergence time is 2, 0, 0,
1 in the 4 respective quadrants of the phase space in Fig.~\ref{phase}.
For the initial state 1,
the game has the same set of convergence times,
except that the order described is permuted.
Hence, the convergence time is 2, 1 and 0
with probabilities $1/4$, $1/4$ and $1/2$ respectively,
yielding the average convergence time of $3/4$.

\begin{figure}
\centerline{\epsfig{figure=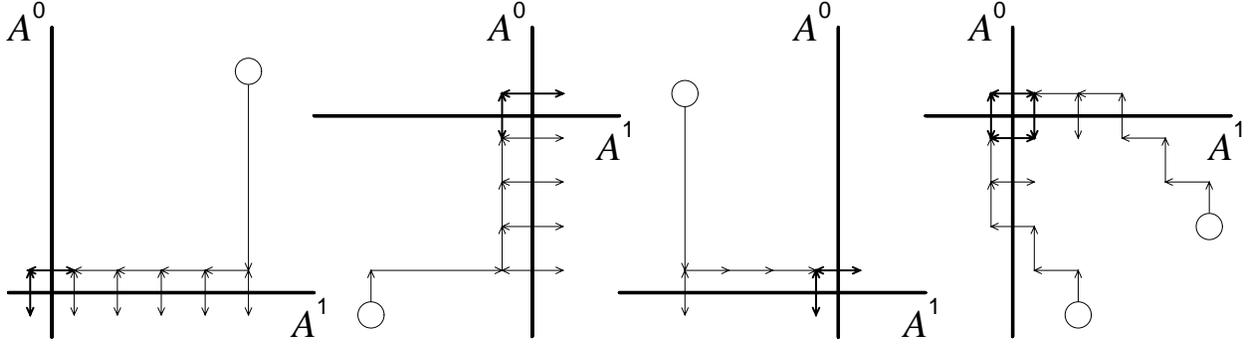,width=1\linewidth}}
\caption{\label{path}
The convergence paths starting from the initial state 0
in the 4 quadrants of the phase space for $m=1$.}
\end{figure}

\begin{figure}
\centerline{\epsfig{figure=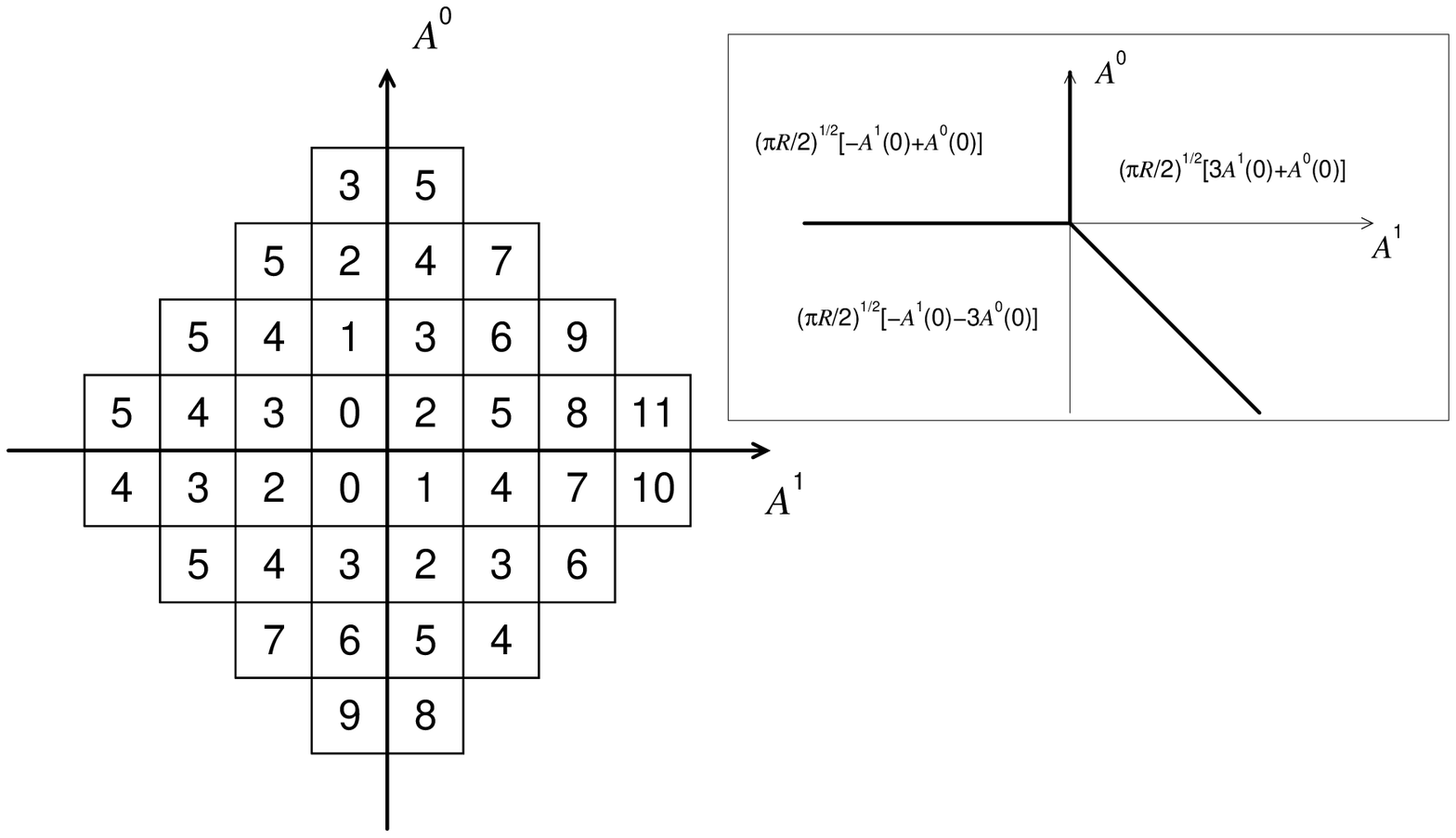,width=1\linewidth}}
\caption{\label{timetile}
The dependence of the convergence time
on the initial position in the phase space for $m=1$,
starting from the initial state 0.
The dimensions of the hypercubes are $\sqrt{2/\pi R}$.
Inset: The 3 regimes of convergence time in the continuum limit.}
\end{figure}

In the scaling regime,
it is convenient to make use of the rectilinear nature of the motion 
in the phase space.
We divide the phase space into hypercubes with dimensions $\sqrt{2/\pi R}$.
Starting from the initial state 0,
the convergence paths are shown in Fig.~\ref{path}.
The convergence time $\tau$ of an initial state 
from inside a hypercube is the number of steps 
it hops between the hypercubes on its way to the attractor,
as shown in Fig.~\ref{timetile}.

In general,
the convergence time is given by the following cases:
(a) $3x+y+2$ for $x\geq 0$ and $y\geq-x-1$,
where $x=\left\lfloor\sqrt{\frac{\pi R}{2}}A^1(0)\right\rfloor$
and $y=\left\lfloor\sqrt{\frac{\pi R}{2}}A^0(0)\right\rfloor$;
(b) $-x-3y-4$ for $y\leq-2$ and $y\leq-x-2$;
(c) $-x+y-1$ for $x\leq-2$ and $y\geq-1$;
(d) $y$ for $x=-1$ and $y\geq0$;
(e) 0 for $x=y=-1$.

The average convergence time is then obtained 
by averging over the Gaussian distribution of the initial $A^\mu(0)$
with mean 0 and variance $1/N$.
When $\rho$ is small,
the initial positions are mainly distributed around the origin, 
reducing the convergence time to that of the multinomial regime.
When $\rho$ is large,
the initial positions are broadly distributed 
among many hypercubes in the phase space,
and one can take a continuum approximation 
as shown in the inset of Fig.~\ref{timetile}.
Thus, the average convergence time is given by
\begin{eqnarray}
        \tau=&&\sqrt{\frac{\pi R}{2N}}
	\Biggl\{\int_0^\infty Dx\int_{-x}^\infty Dy(3x+y)
        +\int_{-\infty}^0Dy\int_{-\infty}^{-y}Dx(-x-3y)
	\nonumber\\
        &&+\int_{-\infty}^0Dx\int_0^\infty Dy(-x+y)\Biggr\},
\end{eqnarray}
where $Dx\equiv dx~e^{-\frac{x^2}{2}}/\sqrt{2\pi}$ is the Gaussian measure.
The result is
\begin{equation}
        \tau=(2+\sqrt{2})\sqrt{\rho}.
\end{equation}
As shown in Fig.~\ref{convtime}, there is an excellent agreement 
between theory and simulations.

The $\rho^{1/2}$ dependence of the convergence time 
can be interpreted as follows.
In the scaling regime,
since the step size in the phase space scales as $1/\sqrt{R}$  
and the initial position of $A^\mu$ has components scaling as $1/\sqrt{N}$,
the convergence time should scale as
$(1/\sqrt{N})/(1/\sqrt{R})\sim\rho^{1/2}$.
This scaling relation remains valid in the kinetic sampling regime 
where $\rho\sim N$,
since kinetic sampling only affects the description of the attractor,
rather than the transient behavior.

\section{\label{sec:wealth}Wealth Spread}

Another system property dependent on the transient
is the distribution of wealth or resources,
especially those among the {\it frozen} agents
(that is, agents who do not switch their strategies at the steady state).
Since the system dynamics reaches a periodic attractor,
they have constant average wealth at the steady state.
Hence any spread in their wealth distribution
is a consequence of the transient dynamics.

To simiplify the analysis,
we only consider the agents who hold identical strategy pairs.
Since they never switch strategies,
and both outputs 1 and 0 have equal occurence at the attractor,
their wealth averaged over a period becomes a constant, 
and their wealth is equal 
to the cumulative payoff of the identical strategies they hold.

In the multinomial regime,
the wealth of agents holding identical strategies $a$ 
is given by Eq.~(\ref{decomp}),
where $k_\mu(t)$ are listed in Table~\ref{m1attstate}.
For $m=1$, the periodic average $\langle\Omega_a\rangle_t$ 
of the cumulative payoffs of strategies 
and their variances $\langle\langle\Omega_a\rangle_t^2\rangle_a$
are listed in Table~\ref{wealthav}.
Thus, the wealth spread $W$ is the variance 
$\langle\langle\Omega_a\rangle_t^2\rangle_a$ of $\langle\Omega_a\rangle_t$,
averaged over the four strategies and the four attractors,
and is equal to $5/8$.

\begin{table}[htp]
\begin{center}
\begin{tabular}{|c|cc|rrrr|} \hline

                             & ~$\xi_a^1$~ & ~$\xi_a^0$~ 
                             & (a) & (b) & (c) & (d)~\\\hline
$\langle\Omega_0\rangle_t$   & -1 & -1
                             & 1 & 0 & -1 & 0\\   
$\langle\Omega_1\rangle_t$   & 1 & -1 
                             & -$\frac{1}{2}$ & $\frac{1}{2}$ 
                             & -$\frac{1}{2}$ & -$\frac{3}{2}$\\ 
$\langle\Omega_2\rangle_t$   & -1 & 1
                             & $\frac{1}{2}$ & -$\frac{1}{2}$ 
                             & $\frac{1}{2}$ & $\frac{3}{2}$\\ 
$\langle\Omega_3\rangle_t$   & 1 & 1
                             & -1 & 0 & 1 & 0\\\hline   
$\langle\langle\Omega_a\rangle^2_t\rangle_a$ & & 
                             & $\frac{5}{8}$ & $\frac{1}{8}$ & $\frac{5}{8}$ 
                             & $\frac{9}{8}$\\\hline 

\end{tabular}
\end{center}
\caption{\label{wealthav}
The variance $\langle\langle\Omega\rangle_t^2\rangle_a$
of the periodic average of wealth of the 4 strategies,
for the 4 attractors of $m=1$.}
\end{table}

In the scaling regime,
the initial position may be located away from the origin of the phase space.
Using the hypercube picture of the transient motion,
we can work out the cumulative payoffs of the strategies
by considering their changes 
when their initial position shift to successive neighboring hypercubes.
The distribution of wealth variance is shown in Fig.~\ref{wealthtile}.
In general, if $x=\left\lfloor\sqrt{\frac{\pi R}{2}}A^1(0)\right\rfloor$
and $y=\left\lfloor\sqrt{\frac{\pi R}{2}}A^0(0)\right\rfloor$,
then the average wealth of the 4 strategies in Table~\ref{wealthav}
are $x+y+1$, $-x+y-1/2$, $x-y+1/2$ and $-x-y-1$ respectively.
This leads to a wealth spread of $x^2+y^2+3x/2+y/2+5/8$.

\begin{figure}
\centerline{\epsfig{figure=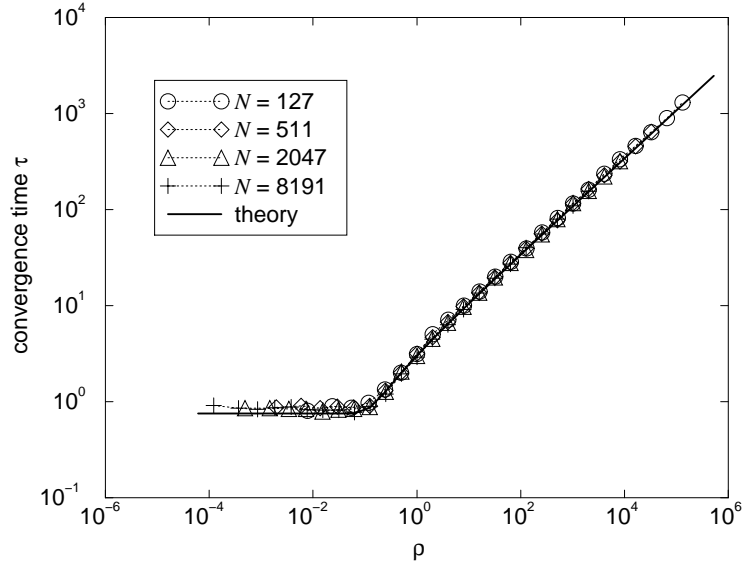,width=0.6\linewidth}}
\caption{\label{convtime}
The dependence of the average convergence time
on the diversity at $m=1$.}
\end{figure}

\begin{figure}
\centerline{\epsfig{figure=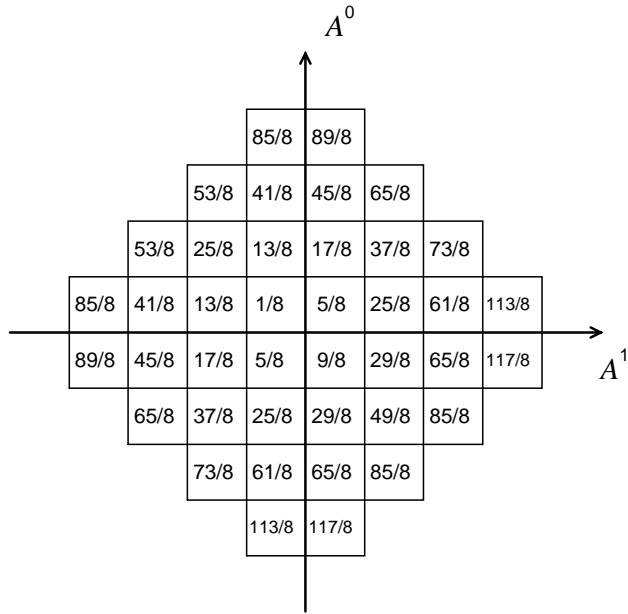,width=0.5\linewidth}}
\caption{\label{wealthtile}
The dependence of the variance of wealth
among the agents holding identical strategies 
on the initial position in the phase space for $m=1$.
The dimensions of the hypercubes are $\sqrt{2/\pi R}$.}
\end{figure}

\begin{figure}
\centerline{\epsfig{figure=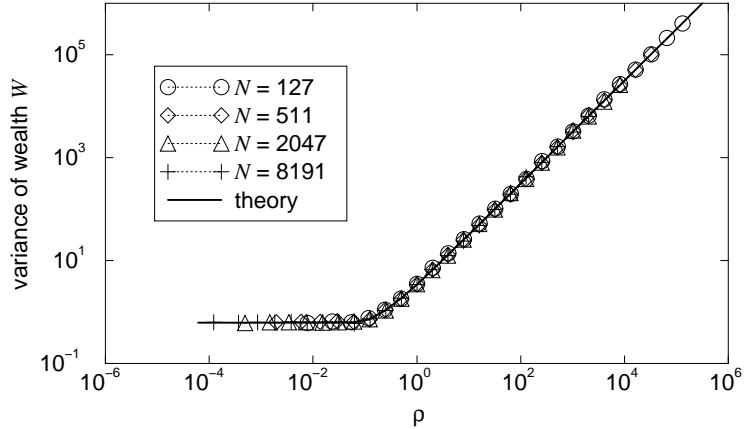,width=0.6\linewidth}}
\caption{\label{wealthvar}
The dependence of the variance of wealth on the diversity
among the agents holding identical strategies for $m=1$.}
\end{figure}

The value of $W$ is then obtained by averaging the wealth spread 
over the Gaussian distribution of
the initial positions in the phase space,
each component $A^\mu(0)$ with mean 0 and variance $1/N$.
When $\rho$ is small,
the initial positions are mainly distributed around the origin,
reducing the wealth spread $W$ to the value at the multinomial regime.
When $\rho$ is large,
the initial positions are broadly distributed 
among many hypercubes in the phase space.
Applying the continuum approximation,
\begin{equation}
        W=\frac{\pi R}{2N}\int Dx\int Dy(x^2+y^2)=\pi\rho.
\end{equation}
The same scaling relation applies to the kinetic sampling regime.
As shown in Fig.~\ref{wealthvar},
agreement between theory and simulations is excellent.
Note that the behavior closely resembles 
that of the convergence time in Fig.~\ref{convtime},
showing that it is a transient behavior.

\section{\label{sec:discuss}Discussions}

We have studied the effects of diversity
in the initial preference of strategies
on a game with adaptive agents competing selfishly for finite resources.
Introducing diversity is both useful 
in modeling agent behavior in economic markets, 
and as a means to improve distributed control. 
We find that it leads to the emergence of a high system efficiency.
We have made use of the small memory sizes $m$ 
to visualize the motion in the phase space.
Scaling of step sizes accounts for the dependence of the efficiency   
on the diversity in the scaling regime ($\rho\sim 1$),
while kinetic sampling effects have to be considered at higher diversity,
yielding theoretical predictions with excellent agreement 
with simulations up to $\rho\sim N$.
However, when diversity increases further, 
waiting effects have to be considered \cite{wong}
and will be discussed in details elsewhere.
The variance of decisions decreases with diversity,
showing that the maladaptive behavior is reduced.
On the other hand,
the convergence time and the wealth spread increases with diversity.

While the present results apply mostly to the cases of small $m$,
qualitative predictions can be made about higher values of $m$.
An extension of Eq.~(\ref{stepsum}) shows that when $\alpha$ increases,   
the step size becomes smaller and smaller in the asymptotic limit.
There is a critical slow down since the convergence time diverges
at $\alpha_c=\pi^{-1}=0.3183$ \cite{wong}.
When $\alpha$ exceeds $\alpha_c$,
the step size vanishes before the system
reaches the attractor near the origin,
so that the state of the system is trapped at locations
with at least some components being nonzero.
The interpretation is that when $\alpha$ is large,
the distribution of strategies become so sparse that
motions in the phase space cannot be achieved by the switching of strategies.
This agrees with the picture of a phase transition
from the symmetric to asymmetric phase
when $\alpha$ increases \cite{symmetry}.
It is interesting to note that the value of $\alpha_c$
is close to the value of 0.3374
obtained by the continuum approximation \cite{challet, bias}
or batch update \cite{coolen} 
using linear payoff functions.

Another extension to general $m$ applies to the symmetric phase
of the exogenous game.
In this case the attractor can be approximated by a hyperpolygon  
enclosing the origin of the phase space.
Using a generating function approach,   
we have computed the variance of decisions,
taking into account the scaling of step sizes and kinetic sampling;
the analysis will be presented elsewhere.   
The results agree qualitatively with simulations  
of both the exogenous and endogenous games,
except for values of $\alpha$ close to $\alpha_c$.
In fact, when $\alpha$ increases,
there is an increasing fraction of samples
in which the attractors are more complex than hyperpolygons.
For example, in the endogenous case,
there is an increasing fraction of attractors
whose periods are no longer $2D$ \cite{caridi}. 
Instead, their periods become multiples of the fundamental period $2D$.  
It is remarkable that the population variance
is not seriously affected by the structural change of the attractor,
probably because the dynamical description of
such long-period attractors have strong overlaps with
those of several distinct attractors of period $2D$.

Besides step payoffs, 
the case of linear payoffs is equally interesting. 
In fact, the latter case has also been considered recently, 
and the variance of decisions is also found 
to decrease with diversity \cite{queenie}. 
There are significant differences between the two cases, though, 
indicating that agents striving to maximize different payoffs 
cause the system to self-organize in different fashions. 
The details will be explained elsewhere.

From the viewpoint of game theory,      
it is natural to consider whether the introduction of diversity
assists the game to reach a Nash equilibrium,
in contrast to the case of the homogeneous initial condition
where maladaptation is prevalent.
It has been verified that Nash equilibria consist of pure strategies
\cite{challet}.
Hence all frozen agents have no incentives to switch their strategies.
In fact, since the dynamics in the attractor is periodic for small $m$,
with states $\pm 1$ appearing once each
in response to each historical string,
the payoffs of all strategies become zero when averaged over a period.
Thus, the Nash equilibrium is approached in the sense that
the fraction of fickle agents decreases with increasing diversity.
In the limit of $\rho\gg N$,
it is probable that only one fickle agent switches strategy 
at each step in the attractor, 
as predicted by Eq.~(\ref{m1ficklek}) for the case $m=1$. 
In this case, agents who switch their decisions
cannot increase their payoffs,
since on switching,
 the minority ones would become losers,
and the majority ones would change the minority side to majority and lose.
(Though the fickle agents are not playing pure strategies,
this argument implies that their payoffs are the same
as if they are doing so.)
Then a Nash equilibrium is reached exactly.  
However, as mentioned previously, 
waiting effects become important in the extremely diverse limit, 
and there are cases that more than one fickle agent 
contribute to a single step in the attractor dynamics, 
and Nash equilibrium cannot be reached.

The combination of scaling and kinetic sampling
in accounting for the steady state properties of the system
illustrates the importance of dynamical considerations
in describing the system behavior,
at least for small values of $m$.
We anticipate that these dynamical effects will play a crucial role
in explaining the system behavior in the entire symmetric phase,
since when $\alpha$ increases,
the state motion in a high dimensional phase space
can easily shift the tail of the cumulative payoff distributions
to the verge of strategy switching,
leading to the sparseness condition
where kinetic sampling effects are relevant. 
Due to their generic nature inherent
in multi-agent systems with dynamical attractors
formed by the collective actions of many adaptive agents,
we expect that these effects are relevant
to minority games with different payoff functions and updating rules,
as well as other multi-agent systems
with adaptive agents competing for limited resources.   

The sensitivity of the steady state to the initial conditions
has implications to adaptation and learning in games.
First, when the MG is used as a model of financial markets,
it shows that the maladaptive behavior is,
to a large extent,
an artifact of the homogeneous initial condition.
In practice,
when agents enter the market with diverse views
on the values of the strategies,   
the corresponding initial condition should be randomized,
and the market efficiency is better than previously believed.
Second, when the MG is used as a model of distributed load balancing,
the present study illustrates the importance
to adopt diverse initial conditions
in order to attain the optimal system efficiency. 
The effect is reminiscent of the dynamics of learning in neural networks, 
in which case an excessive learning rate 
might hinder the convergence to optimum \cite{hkp}.

\begin{acknowledgments}
We thank Peixun Luo, Leihan Tang, Yi-Cheng Zhang, Bing-Hong Wang, 
Jeffrey Chasnov, Andrea de Martino, Esteban Moro, 
Ton Coolen, David Sherrington, Tobias Galla, and Neil Johnson 
for fruitful discussions.
This work is supported by the research grants 
HKUST6153/01P and HKUST6062/02P of the Research Grant Council of Hong Kong.
\end{acknowledgments}

\bibliography{pre_mg}

\begin{thebibliography}{30}
\bibitem[{com({\natexlab{a}})}]{compare1}
\eprint{This result agrees qualitatively with Fig. 11 of \cite{moro} and Fig. 3
  of \cite{coolen}; here we use the data collapse for different $N$ to
  explicitly show that the variance is a function of $\rho$ in the scaling
  regime.}
\bibitem[{com({\natexlab{b}})}]{compare2}
\eprint{This is the regime considered in \cite{heimel} and \cite{coolen}, but
  results differ due to model differences. \cite{heimel} considers batch update
  and linear payoff, whereas we consider on-line update and step payoff. As a
  result, there is a discontinuous transition instead of a scaling relation in
  Fig. 5 of \cite{heimel}.}
\end{thebibliography}

\end{document}